\documentclass{article}
\usepackage{amsmath}
\usepackage{amssymb}
\usepackage{float}
\usepackage{epsfig}
\usepackage{graphicx}
\usepackage{caption}
\usepackage{subcaption}
\usepackage{etoolbox}
\usepackage{amsthm}
\usepackage{ytableau}
\usepackage{tikz}
\usepackage{makecell}
\usepackage[utf8]{inputenc}

\usepackage[a4paper,margin=2.5cm,footskip=0.25in]{geometry}

\newtheorem{theorem}{Theorem}[section]
\newtheorem{corollary}[theorem]{Corollary}
\newtheorem{lemma}[theorem]{Lemma}
\newtheorem{proposition}[theorem]{Proposition}
\newtheorem{remark}[theorem]{Remark}
\theoremstyle{definition}
\newtheorem{Example}[theorem]{Example}%[section]

\counterwithin*{equation}{section}
\counterwithin*{equation}{section}
\newenvironment{prf}{\trivlist \item [\hskip 
\labelsep {\bf Proof:}]\ignorespaces}{\qed \endtrivlist} 
\newcommand{\beq}{\begin{equation}}  
\newcommand{\eeq}{\end{equation}}  
\newcommand{\bea}{\begin{eqnarray}} %% already in Fordy macros 
\newcommand{\eea}{\end{eqnarray}}   %% ditto 
\newcommand{\bear}{\begin{array}}  
\newcommand{\eear}{\end{array}} 

\newcommand\dd{\mathrm{d}}
\newcommand{\D}{{\mathbb D}}

\newcommand{\Z}{{\mathbb Z}}
\newcommand{\C}{{\mathbb C}}
\newcommand{\R}{{\mathbb R}}

\newcommand\ka{{\kappa}}
\newcommand\al{{\alpha}}
\newcommand\be{{\beta}}
\newcommand\ze{{\zeta}}

\newcommand\om{{\omega}}

\newcommand\si{{\sigma}}

\newcommand\tla{\tilde{\lambda}}

\newcommand\sh{{\cal S}}

\newcommand\ve{{\varepsilon}}

\title{Casting more light in the shadows: dual Somos-5 sequences}
\author{J.W.E. Harrow and A.N.W. Hone~\\
School of Mathematics, Statistics \& Actuarial Science~\\
    University of Kent, Canterbury CT2 7NF, UK}
\begin{document}
\maketitle

\begin{abstract} Motivated by the search for an appropriate notion of a cluster superalgebra, incorporating Grassmann variables, 
Ovsienko and Tabachnikov considered the extension of various recurrence relations with the Laurent phenomenon to the ring of dual numbers. 
Furthermore, by iterating recurrences with specific numerical values, some particular well-known integer sequences, such as the Fibonacci sequence, Markoff numbers, and Somos sequences, were shown to 
produce associated ``shadow'' sequences when they were extended to the dual numbers. 
Here we consider the most general version of the Somos-5 recurrence defined over the ring of dual numbers $\D$ with complex 
coefficients, that is the ring  $\C[\varepsilon]$ modulo the relation $\varepsilon^2=0$. 
We present three different ways to present the general solution of the initial value problem for Somos-5 and its shadow part: in analytic form, using the Weierstrass sigma function 
with arguments in $\D$; in terms of the solution of a linear difference equation; and using Hankel determinants constructed from $\D$-valued moments, 
via a connection with a QRT map over the dual numbers. 
\end{abstract}

\section{Introduction}
Supersymmetry is a proposed physical framework in which bosons and fermions can be treated on the same footing. In terms of algebra, this means that one should work with 
a $\Z_2$-graded ring that is a direct sum of even and odd components, and one can then define geometric structures (supergeometry) by working over such a ring. 
One of the simplest examples, 
corresponding to the minimal case of % requirement for 
$N=1$  supersymmetry in physics \cite{bl}, 
is to extend the real numbers $\R$ by two Grassmann variables $\xi_1,\xi_2$, satisfying 
\beq\label{rels}
\xi_j\xi_k + \xi_k\xi_j =0, \qquad j,k=1,2, 
\eeq 
which produces the ring 
\beq\label{Neq1}
(\R\oplus \R \xi_1\xi_2)\oplus (\R\xi_1\oplus\R\xi_2)=R_0\oplus R_1, 
\eeq
whose even part $  R_0=\R\oplus \R \xi_1\xi_2$ contains nilpotent elements, namely the multiples of $\ve=\xi_1\xi_2$. The ring 
(or $\R$-algebra) corresponding to the even part $D=R_0$, that is
$$ 
D=\R\oplus\R\ve \qquad \mathrm{with} \quad \ve^2=0, 
$$ 
is known as the set of dual numbers, and was first introduced by Clifford. 
Replacing $\R$ by $\C$ or another field (or a ring) gives analogues of the dual numbers, which are useful in computer 
algebra (for automatic differentiation) and in algebraic geometry (for defining the tangent space of an algebraic variety). 
For the purposes of this paper, it will be convenient to take the complex numbers as the ambient field, and work with the 
commutative $\C$-algebra of dual numbers $\D=D\otimes \C$  given by 
$$ 
\D=\{ \,x+ y \ve \, | \, x,y\in\C, \, \ve^2=0\,\}, 
$$ 
which is isomorphic to the quotient $\C[\ve]\,/\left<\ve^2\right>$.

Although the algebraic and geometric aspects of supergeometry have been developed for some time, it seems that certain arithmetic aspects of superalgebras have only begun to be explored 
very recently. There are two especially noteworthy examples:  the ``shadows'' of integer sequences   \cite{ovsienko2018dual, ovsienko2021shadow, hone2021casting, veselov}; and the notion of supersymmetric continued fractions associated with the supermodular group OSp($1|2,\Z$) \cite{ovcmodular}, 
namely the supergroup OSp($1|2$) with coefficients in the ring $\Z[\xi_1,\xi_2]$
with two Grassmann variables satisfying (\ref{rels}).
Both of the latter examples have come about as a byproduct  of the search, starting with \cite{ov1}, 
for an appropriate notion of a cluster superalgebra, including mutations of both even and odd cluster variables, 
together with supersymmetric analogues  of related objects like frieze patterns and snake graphs \cite{mgot,  ovshap, moz, musiker}. 

The general philosophy of shadow sequences is explained in 
the paper \cite{ovsienko2021shadow}, where Ovsienko considers %"shadows" of 
integer sequences that are obtained by 
replacing variables $x_n$ in  nonlinear 
recurrence relations by dual variables $X_n = x_n + y_n\ve \in \mathbb{D}$, 
%where $\epsilon^2=0$, 
as well as taking $\D$-valued coefficients %appearing 
in any such  recurrence. % with dual numbers. 
Working in $\D$ (where we can regard  $\ve$ on its own as a single Grassmann variable),  
the resulting recurrence relation for $X_n$ %equation 
can be split into its odd and even parts in terms of powers of $\ve$, with the original recurrence relation is recovered in the even ($\ve^0$) part. The remaining odd ($\ve^1$) part defines the new associated sequence of values $y_n$, satisfying an inhomogeneous linear relation in terms of the $x_n$ and coefficients of the original equation and their new dual parts. 
If we restrict the coefficients to be $\C$-valued, then the sequence $(y_n)$ is referred to as a ``shadow" of the original sequence $(x_n)$: in that case, 
$y_n$ is a solution of a homogeneous linear reurrence, which is just the linearization of the original recurrence around $x_n$. Hence 
a single original sequence $(x_n)$ has multiple shadows, 
belonging to a vector space with the same dimension as the original nonlinear recurrence relations that  it satisfies; but in the case of integer sequences 
$x_n\in\Z$ generated by nonlinear recurrence, the problem of characterizing its integer shadows $y_n\in\Z$ appears to be an interesting arithmetical question. 
As an example, Ovsienko demonstrates that the tetrahedral numbers may be obtained as a shadow of the natural numbers when they are viewed as solutions of the nonlinear relation
\beq\label{affineA1} 
x_{n+1}x_{n-1}=x_n^2-1. 
\eeq 
The latter is one of the simplest examples of a recurrence which exhibits the Laurent phenomenon \cite{fomin2001laurent}, meaning that if the two initial values $x_0,x_1$ are viewed as variables then all of the iterates are Laurent polynomials in these variables with integer coefficients: $x_n\in\Z[x_0^{\pm1},x_1^{\pm 1}]$ for all $n$.\par
Another  example of the Laurent phenomenon is provided %an integer sequence can be given 
by the Somos-5 recurrence, 
\begin{equation}\label{origs5}
    x_{n}x_{n+5}=\alpha x_{n+1}x_{n+4} + \beta x_{n+2}x_{n+3}
\end{equation}
with coefficients $\alpha, \beta$. %This has been proven  to have \cite{fomin2001laurent}
In this case, the Laurent property for (\ref{origs5}) means all iterates of the recurrence are Laurent polynomials in the initial 5 entries $x_0,x_1,x_2,x_3,x_4$ with with coefficients belonging to the polynomial ring $\Z[\alpha, \beta]$, i.e.
\begin{equation}
    x_n \in \mathbb{Z}\left[\alpha, \beta, x_0^{\pm 1}, x_1^{\pm 1}, x_2^{\pm 1}, x_3^{\pm 1}, x_4^{\pm 1}\right], \quad\forall \, n \in \mathbb{Z}.
\end{equation}
It follows that setting all five initial conditions $x_j = 1$ for $j=0,1,2,3,4$ and taking $\al,\be\in\Z$ 
ensures that the whole sequence $(x_n)$ consists of integers for all $n$. The original example considered is when $\al=\be=1$, 
commonly referred to as \textit{the} Somos-5 sequence, in which case  first few terms are given by 
% or Somos(5) and registered on , are given by;
\begin{equation} \label{BaseSomos5Sequence}
    1,1,1,1,1,2,3,5,11,37,83,274,1217,..., 
\end{equation} 
this being the OEIS sequence A006721 \cite{oeis}.  However, there are considerably more choices of initial conditions and coeffcients that give rise to integer sequences, due to the way that 
the recurrence is connected with elliptic curves (see below). \par
Shadow sequences obtained from an analogous recurrence of fourth order, namely Somos-4 given by 
\begin{equation}
    x_nx_{n+4} = \alpha x_{n+1}x_{n+3} + \beta x_{n+2}^2,
\end{equation}
were presented for general initial conditions by one of us in \cite{hone2021casting}. Sequences of Somos type, or 
%is another of a general set of recurrence relations called the Somos-$N$ or, more generally, 
Gale-Robinson sequences, have many structural similarities with Somos-4 and Somos-5 sequences: they inherit the Laurent property via reduction from partial difference equations, 
specifically the octahedron/cube recurrences (a.k.a. the discrete Hirota/Miwa equations)  \cite{fomin2001laurent, mase2016investigation},  which means they can also 
be interpreted as discrete integrable systems \cite{hone2018some}. 

The purpose of this article is to explore how Somos-5 sequences extend to sequences of dual numbers, and to determine explicit expressions for the shadow sequences defined by this change.
In the next section we describe basic properties of Somos-5 and its extension to $\D$. Section 3 presents analytic formulae in terms of Weierstrass functions. 
The fourth section provides a derivation of 
a %  the 
complete set of shadow Somos-5 sequences, given by a combination of analytic and algebraic expressions, based on variation of parameters for a linear difference equation, 
while the fifth section presents Hankel determinant formulae for dual Somos-5 sequences, making use of recent results from   \cite{hone2023family}. Section 6 is concerned with the 
Quispel-Roberts-Thompson (QRT) map associated with Somos-5, and the interpretation of its dual number version as a discrete integrable system, via the construction of a compatible pencil of 
Poisson brackets (a bi-Hamiltonian structure). We end with some brief comments and conclusions.

\section{Somos-5 recurrence and  dual version}

We begin by briefly reviewing some properties of the Somos-5 recurrence, which are summarized in \cite{hone2022heron}. %\eqref{Somos5Recurrence}. 
%Stating the Somos-5 recurrence again:
%\begin{equation} \label{Somos5Recurrence}x_{n}x_{n+5}=\alpha x_{n+1}x_{n+4} + \beta x_{n+2}x_{n+3}\end{equation}This can equivalently be written by defining
%\begin{equation} \label{fnDefinition}    f_n = \frac{x_{n-1}x_{n+1}}{x_n^2},\end{equation}and then the sequence defined by
%\begin{equation} %   f_{n-1}f_n^2f_{n+1}^2f_{n+2} = \alpha f_nf_{n+1}+\beta. %\end{equation} \par
The recurrence (\ref{origs5}) has %can be shown to have 
a 2-invariant  given by
\begin{equation}\label{2invt}
    K_n = \frac{x_nx_{n+4} + \alpha x_{n+2}^2}{x_{n+1}x_{n+3}};
\end{equation}
it is straightforward to verify directly from the recurrence that this satisfies the period 2 condition
%which can be seen to satisfy 
$K_n = K_{n+2}$. 
% by shifting $n$ by 1 in the above, applying the recurrence \eqref{Somos5Recurrence}, and repeating once more. 
This leads immediately to  two independent invariant quantities $I,J$ for the Somos-5 recurrence, which are obtained 
from the product and sum, respectively,  of the odd and even values of this 2-invariant, according to % which we can define by
\begin{align}
    I = K_n + K_{n+1}, \\
    J = \frac{K_nK_{n+1}-\be}{\al}
\end{align}
(where it is assumed that $\al\neq 0$). 
These invariants can both be given explicitly in terms of the coefficients $\al,\be$ and any 5 adjacent iterates, by solving the following pair of homogeneous relations of degree 5 in $x_j$:
\begin{align}
    \begin{split} \label{BaseI}
        x_nx_{n+1}x_{n+2}x_{n+3}x_{n+4}I = &\, x_n^2x_{n+2}x_{n+4}^2 \\ &+ \alpha\left(x_{n+1}^3x_{n+3}x_{n+4}+x_nx_{n+2}^3x_{n+4} + x_nx_{n+1}x_{n+3}^3 \right) \\ &+ \beta x_{n+1}^2x_{n+2}x_{n+3}^2,
    \end{split}\\
    \begin{split} \label{BaseJ}
        x_nx_{n+1}x_{n+2}x_{n+3}x_{n+4}J = &\, x_n^2x_{n+3}^2x_{n+4} + x_nx_{n+1}^2x_{n+4}^2 \\ &+ \alpha\left(x_{n+1}^2x_{n+2}^2x_{n+4} + x_{n}x_{n+2}^2x_{n+3}^2 \right) \\ &+ \beta x_{n+1}x_{n+2}^3x_{n+3}.
    \end{split}  
\end{align}
These relations allow $I,J$ to be simply computed from any set of 5 initial conditions, and then they are both constant along each orbit of (\ref{origs5}). 
Both relations also extend directly to the case of  dual numbers. \par
To extend the recurrence relation (\ref{origs5}) to the dual numbers $\mathbb{D}$ in the most general way possible, we write the dual Somos-5 recurrence as
\begin{equation} \label{dualsomos5}
    X_nX_{n+5} = \alpha X_{n+1}X_{n+4} + \beta X_{n+2}X_{n+3},
\end{equation}
with not only dual number variables $X_n = x_n + y_n\in\ve$, but also  coefficients $\alpha, \beta \in \mathbb{D}$,  
which can be expanded as 
$$\alpha = \alpha^{(0)} + \alpha^{(1)}\ve, \qquad \beta = \beta^{(0)} + \beta^{(1)}\ve.$$ 
In their paper \cite{ovsienko2018dual}, Ovsienko and Tabachnikov consider the specific cases
\begin{align} 
    X_nX_{n+5} &= (1+\alpha^{(1)}\ve) X_{n+1}X_{n+4} + X_{n+2}X_{n+3}, \label{OvsienkoForm1}\\
    X_nX_{n+5} &= X_{n+1}X_{n+4} + (1+\beta^{(1)}\ve) X_{n+2}X_{n+3}, \label{OvsienkoForm2}
\end{align}
in the context of more general Gale-Robinson sequences, showing these produce integer sequences  for $x_0=x_1=x_2=x_3=x_4=1$ and an arbitrary set of 5  integer initial values 
$y_j\in\Z$ for $0\leq j\leq 4$.
\par
We can immediately expand the general case into its odd and even parts, giving the system:
\begin{equation} \label{evenDualSomos5}
     x_{n}x_{n+5} =  \alpha^{(0)} x_{n+1}x_{n+4} + \beta^{(0)} x_{n+2}x_{n+3},
\end{equation}
in the even ($\ve^0$) part and
\begin{align} \label{oddDualSomos5}
    \begin{split}
        x_{n}y_{n+5}+y_{n}x_{n+5} -\alpha^{(0)}\left(x_{n+1}y_{n+4} + y_{n+1}x_{n+4} \right) \,& \\ - \beta^{(0)}\left(x_{n+2}y_{n+3} + y_{n+2}x_{n+3} \right) &= \alpha^{(1)}x_{n+1}x_{n+4} + \beta^{(1)}x_{n+2}x_{n+3}, 
    \end{split}
\end{align}
in the odd ($\ve^1$) part. Clearly the even part recovers the original Somos-5 recurrence \eqref{origs5} 
for $x_n$, with coefficients $\alpha^{(0)}, \beta^{(0)}$, as expected. The odd part gives the inhomogeneous linear relation for the new sequence $(y_n)$. 
The homogeneous version of (\ref{oddDualSomos5}), %this equation, 
corresponding to $\alpha^{(1)}=\beta^{(1)}=0$, gives the shadow sequence $y_n$ in the sense defined by Ovsienko \cite{ovsienko2021shadow}, and is the linearization of the even part. Five linearly independent shadow sequences can then be specified implicitly in terms of 5 independent sets of initial conditions $y_0,y_1,y_2,y_3,y_4$, 
% could then be calculated directly, however 
but in what follows we will derive 5 specific shadow sequences related to the base sequence $x_n$, using the explicit solution to the Somos-5 recurrence. \par
Fomin and Zelevinsky's original proofs of the Laurent property for Somos-5 and other sequences including Somos-4 \cite{fomin2001laurent} are based on treating the initial data as formal variables. The same method carries over directly to \eqref{dualsomos5} with dual numbers, because the dual version has the same form as (\ref{origs5}) and 
$\mathbb{D}$ is a commutative ring. Thus we see that $X_n$ has the Laurent property in its coefficients and initial data, that is 
\begin{equation}
    X_n = \mathbb{Z}\left[\alpha,\beta,X_0^{\pm1},X_1^{\pm1},X_2^{\pm1},X_3^{\pm1},X_4^{\pm1} \right].
\end{equation}
Upon splitting the relation \eqref{dualsomos5} into its even and odd  parts, we can state a more precise version of the Laurent property for each part, 
by making use of the standard reciprocal formula for dual numbers: 
\begin{equation} \label{dualReciprocal}
    (x+y\ve)^{-1} = x^{-1}(1-x^{-1}y\ve).
\end{equation}
Thus we see that the two parts of the system, \eqref{evenDualSomos5} and \eqref{oddDualSomos5}, together have the Laurent property in the sense  that
\begin{align}
    x_n &\in \mathbb{Z}\left[\alpha^{(0)}, \beta^{(0)}, x_0^{\pm 1}, x_1^{\pm 1}, x_2^{\pm 1}, x_3^{\pm 1}, x_4^{\pm 1}\right], \quad\forall \, n \in \mathbb{Z}. \\
    y_n &\in \mathbb{Z}\left[\alpha^{(0)}, \beta^{(0)}, \alpha^{(1)}, \beta^{(1)}, x_0^{\pm 1}, x_1^{\pm 1}, x_2^{\pm 1}, x_3^{\pm 1}, x_4^{\pm 1}, y_0, y_1, y_2, y_3, y_4\right], \quad\forall \, n \in \mathbb{Z}.
\end{align} 
Also note that to iterate the dual Somos-5 for numerical values at any particular step $n$, we require $x_n\neq 0$ in order to determine $x_{n+5}$.  
This means that if we take the 5 dual initial data to be units, 
i.e. $X_0,X_1,X_2,X_3,X_4 \in \mathbb{D}^*$ where $\mathbb{D}^* = \{x+y\ve\in \mathbb{D}\, |\, x\neq0\}$, 
then the Laurent property ensures that the whole orbit $(X_n)_{n\in\Z}$  can be defined by evaluating suitable Laurent polynomials in these 5 initial values.\par
The conserved quantities likewise carry over to the dual system, as an analogous 2-invariant $K_n\in\D$ can be defined with exactly the same formula \eqref{2invt} 
but replacing each $x_j\to X_j$ and with $\al\in\D$. %and shown to be invariant after $n \rightarrow n+2$ using the recurrence. 
Thus dual invariants (first integrals) $I=I^{(0)}+I^{(1)\ve}$ and $J=J^{(0)}+J^{(1)}\ve$ can be found. %The simplest way of 
To find these explicitly it is best to  simply take the original homogeneous relations %expressions in terms of the regular coefficients and initial data,
\eqref{BaseI} and \eqref{BaseJ}, and change all terms to their dual counterparts, $x_0 \rightarrow X_0$ etc. 
%, as these expressions are found in the same process in both cases. Again the new dual invariants can be split into their odd and even parts. \par

In what follows, the first integral $J\in\D$ will play a central role, so we consider it first.  
For $J$, the even part of the dual version of \eqref{BaseJ} yields precisely the same formula as  \eqref{BaseJ} does for $J^{(0)}$, as expected, 
but with the parameters replaced by $\al^{(0)},\be^{(0)}$: 
\beq\label{j0exp} %{DualJDefinition}
J^{(0)} = \frac{x_n^2x_{n+3}^2x_{n+4}+x_nx_{n+1}^2x_{n+4}^2 + \al^{(0)} (x_{n+1}^2x_{n+2}^2x_{n+4}+x_nx_{n+2}^2x_{n+3}^2) 
+\be^{(0)}x_{n+1}x_{n+2}^3x_{n+3}}{x_nx_{n+1}x_{n+2}x_{n+3}x_{n+4}} 
\eeq 
As it stands, dualizing \eqref{BaseJ} and taking the odd part gives a relation for $J^{(1)}$ which includes $J^{(0)}$ as well. After substituting for $J^{(0)}$, we can get an expression for $J^{(1)}$ in terms of the initial data and coefficients alone:
\begin{equation} \label{DualJDefinition}
    J^{(1)}= \frac{D_n - \sum^4_{j=0}C^{(j)}_nx_{n+j}^{-1}y_{n+j}}{x_nx_{n+1}x_{n+2}x_{n+3}x_{n+4}},
\end{equation}
where
\begin{align}\label{Cdefs} 
    C_n^{(0)}&= \alpha^{(0)}x_{n+1}^2x_{n+2}^2x_{n+4}+\beta^{(0)}x_{n+1}x_{n+2}^3x_{n+3}-x_n^2x_{n+3}^2x_{n+4}, \\
    C_n^{(1)}&= \alpha^{(0)}\left(x_nx_{n+2}^2x_{n+3}^2-x_{n+1}^2x_{n+2}^2x_{n+4} \right)+x_{n}^2x_{n+3}^2x_{n+4}-x_nx_{n+1}^2x_{n+4}^2, \\
    \begin{split}
        C_n^{(2)}&= x_n^2x_{n+3}^2x_{n+4} + x_nx_{n+1}^2x_{n+4}^2
        - \alpha^{(0)}\left(x_{n+1}^2x_{n+2}^2x_{n+4}+x_nx_{n+2}^2x_{n+3}^2 \right)\\ \,& \quad - 2\beta^{(0)}x_{n+1}x_{n+2}^3x_{n+3},
    \end{split} \\
    C_n^{(3)}&= \alpha^{(0)}\left(x_{n+1}^2x_{n+2}^2x_{n+4}-x_nx_{n+2}^2x_{n+3}^2 \right)+x_nx_{n+1}^2x_{n+4}^2-x_{n}^2x_{n+3}^2x_{n+4}, \\
     C_n^{(4)}&= \alpha^{(0)}x_{n}x_{n+2}^2x_{n+3}^2+\beta^{(0)}x_{n+1}x_{n+2}^3x_{n+3}-x_nx_{n+1}^2x_{n+4}^2,\\
     D_n &= \alpha^{(1)}\left(x^2_{n+1}x^2_{n+2}x_{n+4}+x_nx_{n+2}^2x_{n+4} + x_{n}x_{n+2}^2x_{n+3}^2 \right) + \beta^{(1)}x_{n+1}x_{n+2}^3x_{n+3}.
\end{align}
As we might expect, there are many symmetries in this expression. This would be even more apparent if we shifted $n$ down by two to see how the $n-j$ and $n+j$ terms for $j=0,1,2$ balance each other in many expressions. We will use this expression later when examining one of the linearly independent shadow sequences of Somos-5 and note some of its other properties. \par
For completeness, we also include the form of the second dual invariant, $I\in\D$. Similarly the even part gives $I^{(0)}$ by the same expression \eqref{BaseI} for the original $I$. 
The odd part of $I$ can then be found from the odd component, to yield 
\begin{equation} \label{DualIDefinition}
    I^{(1)}= \frac{\bar{B}_n - \sum^4_{j=0}\bar{A}^{(j)}_nx_{n+j}^{-1}y_{n+j}}{x_nx_{n+1}x_{n+2}x_{n+3}x_{n+4}},
\end{equation}
where
\begin{align}
    \bar{A}^{(0)}_n &= \alpha^{(0)}x_{n+1}^3x_{n+3}x_{n+4}+\beta^{(0)}x_{n+1}^2x_{n+2}x_{n+3}^2-x_n^2x_{n+2}x_{n+4}^2, \\
    \bar{A}^{(1)}_n &= x_n^2x_{n+2}x_{n+4}^2 + \alpha^{(0)}\left[x_nx_{n+2}^3x_{n+4}-2x_{n+1}^3x_{n+3}x_{n+4} \right]-\beta^{(0)}x_{n+1}^2x_{n+2}x_{n+3}^2  , \\
    \bar{A}^{(2)}_n &= \alpha^{(0)}\left[ x_{n+1}^3x_{n+3}x_{n+4}-2x_nx_{n+2}^3x_{n+4}+x_nx_{n+1}x_{n+3}^3\right]  , \\
    \bar{A}^{(3)}_n &= x_n^2x_{n+2}x_{n+4}^2+\alpha^{(0)}\left[x_nx_{n+2}^3x_{n+4}-2x_nx_{n+1}x_{n+3}^3 \right]-\beta^{(0)}x_{n+1}^2x_{n+2}x_{n+3}^2 , \\
    \bar{A}^{(4)}_n &= \alpha^{(0)}x_{n}x_{n+1}x_{n+3}^3+\beta^{(0)}x_{n+1}^2x_{n+2}x_{n+3}^2-x_n^2x_{n+2}x_{n+4}^2, \\
    \bar{B}_n &= \alpha^{(1)}\left[x_{n+1}^3x_{n+3}x_{n+4}+x_nx_{n+2}^3x_{n+4}+x_{n}x_{n+1}x_{n+3}^3\right] + \beta^{(1)}x_{n+1}^2x_{n+2}x_{n+3}^2. 
\end{align}
The quantity $I$ is considerably less useful in studying the Somos-5 shadow solutions, 
essentially %as with $I$ and the original Somos-5 sequences. This is 
because the  quantity $J$ leads to a  connection with elliptic curves, which leads to explicit analytic solutions of the recurrence in terms of Weierstrass functions, as 
described in the next section. 
%\par In the next section, we will review an analytic solution to the regular Somos-5 recurrence, discuss how this extends to the dual system, and how the shadow sequences arise from this.

\section{Analytic solution of dual Somos-5}

An explicit analytic solution for the Somos-5 recurrence was given in \cite{hone2006sigma}, 
via the relation with elliptic curves and associated functions. 
We will state one version of this solution here, following the notation used in \cite{hone2022heron}, and subsequently show how to extend it to 
dual Somos-5 sequences. %associated properties that will be relevant later. \par

\begin{theorem} \label{SomosSolutionTheorem}
    The general solution of the initial value problem for the Somos-5 recurrence  \eqref{origs5} over $\mathbb{C}$ is 
    \begin{equation} \label{somos5Solution}
        x_n = A_{\pm}B_{\pm}^{\left\lfloor\frac{n}{2}\right\rfloor}\mu^{\left\lfloor\frac{n}{2}\right\rfloor^2}\sigma(n\kappa+z_0) 
    \end{equation}
(where the subscripts $+/-$ apply for even and odd $n$ respectively), 
given in terms of the 
Weierstrass sigma function $\sigma(z)=\sigma(z;g_2,g_3)$ associated with the elliptic curve 
    \begin{equation} \label{Somos5Curve} %{SomosSolutionTheorem}
       \mathrm{y}^2 = 4\mathrm{x}^3-g_2\mathrm{x}-g_3
    \end{equation}
with the parameters $g_2,g_3,\mu$ and $\ka$ appearing in \eqref{somos5Solution} being 
explicitly determined from the coefficients $\alpha, \beta$ of \eqref{origs5} and the conserved quantity $J$ defined by \eqref{BaseJ}, 
via the formulae 
\begin{equation}\label{g23} 
        g_2 = 12\tla-2J, \quad\quad\quad g_3 = 4\tla^3-g_2\tla-\Tilde{\mu}^2.
\end{equation}
\begin{align} \label{SolutionConstants}
        \Tilde{\mu} = (\beta+\alpha J)^{\frac{1}{4}}, &&
        \tla = \frac{1}{3\Tilde{\mu}^2}\left(\frac{J^2}{4}+\alpha\right), &&
        \mu = \frac{\Tilde{\mu}}{\sigma(2\kappa)} = -\sigma(\kappa)^{-4}.
    \end{align}
%(\ref{g23}) and the formulae    
The arbitrary parameter $z_0$ is determined from the initial data, 
while the remaining  constants $A_{\pm}, B_{\pm}$ (also related to the initial data) are arbitrary up to the constraint
\begin{equation} \label{Bconstraint}
        B_+ = -\mu^{-1}B_- =\sigma(\kappa)^4B_-.
\end{equation}
\end{theorem}
The solution \eqref{somos5Solution} % {SolutionConstants}
corresponds to a sequence of points $P_0+nP$ along the curve \eqref{Somos5Curve}, for an arbitrary initial point 
$P_0=(\wp(z_0),\wp'(z_0))$  translated by $P= (\wp(\kappa),\wp'(\kappa))= (\tla,\Tilde{\mu})$.
The proof in \cite{hone2006sigma} makes use of the fact that if $x_n$ is a solution of  \eqref{origs5} then the 
sequence of ratios 
\begin{equation} \label{oldVolterraSolution}
    w_n = \frac{x_{n+2}x_{n-1}}{x_{n+1}x_n},
\end{equation}
satisfies a nonlinear recurrence relation of second order, namely
\begin{equation} \label{OldVolterraMap}
    w_{n-1}w_nw_{n+1}= \alpha w_n + \beta, 
\end{equation}
which is a particular example of a QRT map in the plane. 
The conserved quantity $J$ for Somos-5 can be rewritten in terms of two adjacent ratios \eqref{oldVolterraSolution}, as the 
expression 
\begin{equation}\label{biqJ}
    J = w_{n-1}+w_n+\alpha\left(\frac{1}{w_{n-1}}+\frac{1}{w_n} \right)+\frac{\beta}{w_{n-1}w_n}.
\end{equation}
The explicit solution is found %explicitly 
by associating the curve \eqref{Somos5Curve} with the recurrence, where the  coefficients are given in terms of the sigma function by 
\begin{align}
\alpha = \frac{\sigma(3\kappa)}{\sigma(\kappa)^9}, && \beta = -\frac{\sigma(4\kappa)}{\sigma(2\kappa)\sigma(\kappa)^{12}}.   
\end{align}
For fixed initial conditions, the parameters $z_0, \kappa$ can be computed from the elliptic integrals
\beq\label{ellint}
    z_0 = \pm\int^{\mathrm{x}_0}_{\infty}\frac{\dd\mathrm{x}}{\mathrm{y}}, 
\qquad \kappa = \pm\int^{\tla}_{\infty}\frac{\dd\mathrm{x}}{\mathrm{y}},
\eeq 
where $\mathrm{x}_0$ and a consistent relative choice of signs above must be determined from  
%signs determined 
%by consistency with 
$$\mathrm{x}_0 = \tla+\frac{\tilde{\mu}^2}{w_{-1}+w_0-J} , 
\qquad 
\Tilde{\mu} = \wp'(\kappa), \qquad \wp'(\kappa)\wp'(z_0)=(\mathrm{x}_0-\lambda)(w_{-1}-w_0),$$ 
while the coefficients $A_{\pm}, B_{\pm}$ can be found from the initial data. 
To see that the number of parameters in the analytic solution match the initial value problem, note that  
the recurrence (\ref{origs5}) requires 5 pieces of initial data to be specified, $x_0,x_1,x_2,x_3,x_4$, say, together with 2 coefficients $\alpha, \beta$, making a 
total of 7 parameters, while the analytic solution  \eqref{somos5Solution}  is completely specified by choosing the 7 quantities $A_+,A_-,B_-,z_0,\kappa,g_2,g_3$. 
(Note that $B_+$ is fixed by the other data by the constraint \eqref{Bconstraint}, 
while $\mu$ is determined from \eqref{SolutionConstants}.)\par
To extend this to the dual system, we note the standard identity 
\begin{equation} \label{dualExpansion}
    \Phi(X)=\Phi(x)+\Phi'(x)y\ve, 
\end{equation}
for any differentiable function $\Phi$ and $X=x+y\ve \in \mathbb{D}$. 
% we can use a simple expansion and $\epsilon^2=0$ to see that the value of $\Phi$ at $X$ can be given by
This can be used to give solutions to the dual system \eqref{dualsomos5} in terms of the analytic solution \eqref{somos5Solution} and derivatives.
\begin{proposition} \label{DualSomosSolutionProp}
    Given dual number parameters $A_{\pm}=A^{(0)}_{\pm}+A^{(1)}_{\pm}\ve,\, B_{-}=B^{(0)}_{-}+B^{(1)}_{-}\ve,\, Z_{0}=z^{(0)}+z^{(1)}\ve,\, K=\kappa^{(0)}+\kappa^{(1)}\ve,\, g_{2}=g^{(0)}_2+g_2^{(1)}\ve,\, g_{3}=g^{(0)}_3+g_3^{(1)}\ve$, with $A_\pm, B_-,\si(K) \in\D^*$, and letting $ \mu =- \sigma(K)^{-4}$, $ B_+ = -\mu^{-1}B_-$, 
the sequence 
    \begin{equation} \label{Dualsomos5Solution}
        X_n = A_{\pm}B_{\pm}^{\lfloor\frac{n}{2}\rfloor}\mu^{\lfloor\frac{n}{2}\rfloor^2}\sigma(nK+Z_0), 
\qquad n\in\Z    \end{equation}
%    where $ B_+ = \sigma(K)^4B_-$, 
satisfies the dual Somos-5 recurrence \eqref{dualsomos5} with coefficients
    \begin{align}
        \alpha = \alpha^{(0)} + \alpha^{(1)}\ve = \frac{\sigma(3K)}{\sigma(K)^9}, && \beta =\beta^{(0)} + \beta^{(1)}\ve= -\frac{\sigma(4K)}{\sigma(2K)\sigma(K)^{12}}.  
    \end{align}
   % provided $A_{\pm},B_- \in \mathbb{D}^*$ and $\sigma(Z)=\sigma(Z;G_2,G_3)\in \mathbb{D}^*$. 
In terms of even/odd components, this may be written as 
\small
    \begin{align} \label{ExpandedDualSolution} % {Dualsomos5Solution}
 X_n = x_n + x_n &\, \left(  \frac{A_{\pm}^{(1)}}{A_{\pm}^{(0)}}+\left\lfloor\frac{n}{2}\right\rfloor\frac{B_\pm^{(1)}}{B_\pm^{(0)}}  +z_0^{(1)}\zeta(z_0^{(0)}+n\kappa^{(0)}) 
+\left(\kappa^{(1)}\partial_{\kappa^{(0)}}+g_2^{(1)}\partial_{g_2^{(0)}}+g_3^{(1)}\partial_{g_3^{(0)}}\right)\log x_n \right)\ve
       % \begin{split}
   %         X_n = x_n + x_n &\, \left(  \frac{A_{\pm}^{(1)}}{A_{\pm}^{(0)}}+\left\lfloor\frac{n}{2}\right\rfloor\frac{B_\pm^{(1)}}{B_\pm^{(0)}}  +z_0^{(1)}\zeta(n\kappa^{(0)}+z_0^{(0)})\right. \\ 
      %       &\left.\quad +\left(\kappa^{(1)}\partial_{\kappa^{(0)}}+g_2^{(1)}\partial_{g_2^{(0)}}+g_3^{(1)}\partial_{g_3^{(0)}}\right)\log(x_n)\right)\ve
       % \end{split}
    \end{align}
\normalsize    
where $\partial$ denotes a partial derivative, 
$\zeta(z)=\zeta(z;g_2^{(0)},g_3^{(0)})$ is the Weierstrass zeta function, 
and $x_n$ is the right-hand side of \eqref{Dualsomos5Solution} with all parameters replaced by their even components, % with superscript $(0)$, %counterparts, ie 
i.e. $A_+ \rightarrow A_+^{(0)}$, etc.% Here  and .
\end{proposition}
\begin{prf} %\begin{proof}
This is analogous to the proof of the analytic solution of the dual Somos-4 recurrence in \cite{hone2021casting}. We can make use of part of the original proof of Theorem 
\ref{SomosSolutionTheorem} for the regular Somos-5 recurrence in \cite{hone2006sigma}: in one direction, the fact that the analytic expression satisfies 
the recurrence relies only the three-term relation for the sigma function. The sigma function $\si(z;g_2,g_3)$ is an analytic function of the argument $z\in\C$, 
and of the parameters $g_2,g_3$, and so the three-term relation holds as an identity of formal series, which is still valid when $z,g_2,g_3$ are replaced by elements 
of the commutative ring $\D$. 
% the properties of the sigma function to associate the recurrence with an ellipt ic curve,  the right-hand side of \eqref{Dualsomos5Solution}
%all steps of which will carry over to $\mathbb{D}$. The proof could then follow equivalently for $\sigma(Z)\in \mathbb{D}^*$ so that divisions by this quantity may always occur, and for $A_{\pm}, B_-\in\mathbb{D}^*$ so that $x_n\neq 0$ for all $n$. \par
As for the even and odd parts, 
we can expand  the right-hand side of \eqref{Dualsomos5Solution} 
 into its even and odd parts, $X_n = x_n + y_n\ve$, by writing 
$$ 
A\pm=A_\pm^{(0)} \left(1+\frac{A_\pm^{(1)}}{A_\pm^{(0)}}\,\ve \right), \quad 
B_{\pm}^{\lfloor\frac{n}{2}\rfloor}=(B_{\pm}^{(0)})^{\lfloor\frac{n}{2}\rfloor}\left(1+\frac{B_\pm^{(1)}}{B_\pm^{(0)}}\,\ve \right)^{\left\lfloor\frac{n}{2}\right\rfloor}=
(B_{\pm}^{(0)})^{\lfloor\frac{n}{2}\rfloor}\left(1+\left\lfloor\frac{n}{2}\right\rfloor\frac{B_\pm^{(1)}}{B_\pm^{(0)}}\,\ve \right),
$$ 
where we used the binomial theorem, 
and also 
$$ 
\begin{array}{rcl}
\si (Z_0+nK;g_2,g_3)& =& \si(z_0^{(0)}+nK;g_2,g_3)\Big(1+z_0^{(1)}\zeta(z_0+nK;g_2,g_3)\ve\Big) \\
& =& 
\si(z_0^{(0)}+nK;g_2,g_3)\Big(1+z_0^{(1)}\zeta(z_0+n\ka^{(0)};g_2^{(0)},g_3^{(0)})\ve\Big)
\end{array} 
$$
by \eqref{dualExpansion}. 
Upon multiplying out and keeping only terms of order zero and one in $\ve$, this  yields 
$$ 
X_n = 
A_\pm^{(0)}(B_{\pm}^{(0)})^{\lfloor\frac{n}{2}\rfloor}\mu^{\lfloor\frac{n}{2}\rfloor^2}\si(z_0^{(0)}+nK;g_2,g_3)
\left(1+\Big(
\frac{A_\pm^{(1)}}{A_\pm^{(0)}} 
+\left\lfloor\frac{n}{2}\right\rfloor\frac{B_\pm^{(1)}}{B_\pm^{(0)}} 
+ z_0^{(1)}\zeta(z_0+n\ka^{(0)};g_2^{(0)},g_3^{(0)})
\Big)\,\ve\right),
$$ 
whose 
even ($\ve^0$) coefficient $x_n$  
%As expected, $x_n$ 
just corresponds to the solution of the regular Somos-5 
recurrence with all parameters being the even parts of the dual ones, while at order $\ve^1$ above one can see the first three terms
that make up the expression for $y_n$  in  
\eqref{ExpandedDualSolution}. The remaining terms at order  $\ve^1$, 
involving a linear combination of the partial derivatives  
$\partial_{\kappa^{(0)}},\partial_{g_2^{(0)}},\partial_{g_3^{(0)}}$ applied to $\log x_n$, follow by considering the 
analytic dependence of the factor $\mu^{\lfloor\frac{n}{2}\rfloor^2}\si(z_0^{(0)}+nK;g_2,g_3)$ 
on the parameters $K,g_2,g_3\in\D$ and applying  \eqref{dualExpansion} to each of these variables in turn. 
\end{prf} 
%\end{proof}

As it stands, the preceding result is not a complete analogue of Theorem \ref{SomosSolutionTheorem}: 
while \eqref{Dualsomos5Solution} provides a solution of the dual Somos-5 recurrence, depending on 7 dual parameters, we have not 
shown that it solves the general initial value problem for \eqref{dualsomos5}. To do this, we would need to have an analogue of the 
elliptic integrals (\ref{ellint}) in order to reconstruct the parameters $Z_0,K\in\D$ from $\al,\be$ and the initial data. This would seem to require 
a proper theory of elliptic curves and integrals over the dual numbers, which seems to be lacking. Nevertheless, the analytic formulae 
appearing in Proposition \ref{DualSomosSolutionProp} will be useful in what follows, for the construction of the shadow Somos-5 sequences.

\section{Somos-5 shadows}
Returning to the recurrence  \eqref{oddDualSomos5} for $y_n$, we noted previously that the case $\alpha^{(1)}=0=\beta^{(1)}$, 
that is 
\begin{equation} \label{linearizedOddSomos5}
     x_{n}y_{n+5}+y_{n}x_{n+5} -\alpha^{(0)}\left(x_{n+1}y_{n+4} + y_{n+1}x_{n+4} \right) - \beta^{(0)}\left(x_{n+2}y_{n+3} + y_{n+2}x_{n+3} \right) = 0,
\end{equation} 
corresponds to the 
linearization of \eqref{evenDualSomos5}, whose solutions are the shadow sequences in the sense of  \cite{ovsienko2021shadow}. 
The above homogeneous linear equation for $y_n$ is of order five,
the same as the order of 
%we note that we have the solution of 
\eqref{evenDualSomos5}, hence the shadow Somos-5 sequences form a vector space of dimension 5.  
In particular, we can consider differentiating the explicit solution  \eqref{somos5Solution} with respect to suitable parameters, in 
order to obtain 5 linearly independent solutions of the linearized equation \eqref{linearizedOddSomos5}. 
\begin{lemma}\label{shadlem} 
For fixed coefficients $\alpha^{(0)},\beta^{(0)}$, 
the Somos-5 shadow equation \eqref{linearizedOddSomos5} has 5 linearly independent solutions, which can be chosen as follows:
    \begin{align} 
        y_n^{(i)} &= \begin{cases}
     0 \quad&\text{for} \,\, n \,\, \text{odd},\\
     x_n \quad &\text{for} \,\, n \,\, \text{even},\\
    \end{cases} \label{Shadow1} \\
    y_n^{(ii)} &= \begin{cases}
     x_n \quad&\text{for} \,\, n \,\, \text{odd},\\
     0 \quad &\text{for} \,\, n \,\, \text{even},\\
    \end{cases} \label{Shadow2}\\
    y_n^{(iii)}&= n\,x_n \label{Shadow3} \\
    y_n^{(iv)}&=\zeta(z_0^{(0)}+n\kappa^{(0)})\, x_n \label{Shadow4}\\
    y_n^{(v)}&=x_n\partial_{J^{(0)}}\log(x_n)= x_n\left(\frac{\dd\kappa^{(0)}}{\dd J^{(0)}}\partial_{\kappa^{(0)}}+\frac{\dd g_2^{(0)}}{\dd J^{(0)}}\partial_{g_2^{(0)}}+\frac{\dd g_3^{(0)}}{\dd J^{(0)}}\partial_{g_3^{(0)}} \right)\log x_n \label{Shadow5}
    \end{align}
\end{lemma}
\begin{prf}
For fixed $\alpha,\beta\in\C$, the solution \eqref{somos5Solution} depends on the 5 complex parameters
$A_+,A_-,B_-,z_0,J$,  which can be chosen arbitrarily, so we can obtain independent solutions 
of the linearized equation by differentiating with respect to each of these in turn. 
In the context of the dual equation, we wish to consider each of these  parameters as the even part of 
a corresponding dual number, so that they should acquire a superscript (0), but 
for the purposes of this proof the superscripts are omitted. 
The parameters $A_{\pm}$ appear alternately in even/odd terms, and the derivatives with respect to these scaling parameters are proportional to $x_n$ for each choice of parity of $n$, hence produce the alternating forms of \eqref{Shadow1} and \eqref{Shadow2} above. 
Using the constraint of \eqref{Bconstraint} to rewrite $B_+$ in terms of $B_-$, $B_-$ appears in both odd and even terms, and the derivative with respect to $B_-$ up to scaling brings down a $\left\lfloor\frac{n}{2}\right\rfloor$ factor on each term, giving $\left\lfloor\frac{n}{2}\right\rfloor x_n$. Then we can set $y_n^{(iii)}=y_n^{(ii)}+2\left\lfloor\frac{n}{2}\right\rfloor x_n$ to obtain the simpler  form of \eqref{Shadow3}. 
The fourth solution, as in \eqref{Shadow4}, contains the Weierstrass zeta function $\zeta(z)=\zeta(z;g_2,g_3)$ for the same $g_2,g_3$ as in the solution, which comes from taking the derivative of $\sigma(z_0+n\kappa)$ and using the definition
$$
        \zeta(z;g_2,g_3) = \frac{\sigma'(z;g_2,g_3)}{\sigma(z;g_2,g_3)}.
$$
    The final linearly independent solution, $y_n^{(v)}$, is found using the derivative with respect to the quantity $J$, the conserved quantity of the Somos-5 sequence, defined by 
\eqref{BaseJ}, since the parameters $\ka,g_2,g_3$ appearing in the formula   \eqref{somos5Solution} all depend on this quantity. The dependence on $J$ is not straightforward, 
since it is determined by variations of the sigma function both with respect to its argument as a quasiperiodic function on the torus $\C/\Lambda$, and with respect to the modular 
parameters $g_2,g_3$ which determine the period lattice $\Lambda$ of the elliptic curve    \eqref{Somos5Curve}.
Hence \eqref{Shadow5} is best left in the form of a sum of  logarithmic derivatives of $x_n$ with respect to the quantities $\kappa,g_2,g_3$. 
% depend on this quantity and these derivatives are not to compute explicitly.
\end{prf}

There are several different ways to construct these explicit shadow solutions. On the one hand, notice that the result of  Lemma \ref{shadlem}  
can be viewed as a corollary of Proposition \ref{Dualsomos5Solution}, since when $\al^{(1)}=\be^{(1)}=0$ the odd part of the  
solution  \eqref{ExpandedDualSolution} satisfies the homogeneous equation \eqref{linearizedOddSomos5}: the freedom of choice in the pair of 
parameters $A_{\pm}^{(1)}$ gives linear combinations of $y_n^{(i)}$ and $y_n^{(ii)}$; 
terms involving the parameters $B_{\pm}^{(1)}$ (of which only one can be chosen 
freely) results in including multiples of  $y_n^{(iii)}$; the parameter $z_{\pm}^{(1)}$ produces linear 
combinations of  $y_n^{(iv)}$; and the last three logarithmic derivatives correspond to the inclusion of the 
fifth linearly independent solution  $y_n^{(v)}$.  
On the other hand, it is also easy to verify that $y_n^{(i)},y_n^{(ii)},y_n^{(iii)}$ are shadow solutions by direct substitution. %induction. %, for example:
\begin{Example}
    The solution $ y_n^{(iii)}= n x_n $ can be shown  to satisfy the linear homogeneous recurrence 
as follows: % \par
Substituting $y_n=nx_n$ into the left-hand side of \eqref{linearizedOddSomos5} produces 
\small
   % \begin{align}%(2n+5)
$$ 
        (n+5)x_nx_{n+5}+nx_nx_{n+5} -\alpha^{(0)} \big( (n+4)x_{n+1}x_{n+4}+(n+1) x_{n+1}x_{n+4}\big)  
-\beta^{(0)} \big((n+3)x_{n+2}x_{n+3}+(n+3)x_{n+2}x_{n+3}\big) , 
$$ 
\normalsize 
which is equal to 
$$ 
(2n+5)\Big(x_{n}x_{n+5}-\al^{(0)} \,x_{n+1}x_{n+4}-\be^{(0)}\,x_{n+2}x_{n+3}\Big) = 0, 
 $$
\normalsize
 since $x_n$ is a solution of the original Somos-5 recurrence \eqref{origs5} with $\al\to \al^{(0)}$,  $\be\to \be^{(0)}$.
\end{Example}
Similar direct substitutions verify that $y_n^{(i)}$ and $y_n^{(ii)}$ are shadow solutions, and even more straightforward to check 
is their sum $y_n = y_n^{(i)} + y_n^{(ii)}= x_n$.  Compared with the analogous results for Somos-4 in  \cite{hone2021casting}, where the original sequence is also a shadow,  
the fact that the splitting of the original solution $x_n$ into its odd and even index components  are separate shadow solutions 
for Somos-5 is not unexpected: it is a consequence of the fact that 
the nonlinear recurrence admits 
a symmetry whereby the odd and even terms of the sequence can be scaled independently. \par
The solution $y_n^{(iv)}$ is more complicated, being expressed in terms of  the Weierstrass $\zeta$ function associated with the elliptic curve \eqref{Somos5Curve}. 
To obtain algebraic relations for this shadow, we note that from the solution \eqref{somos5Solution}, after shifting the index, the ratios $w_n$ in    \eqref{oldVolterraSolution} are given by 
\begin{equation}\label{zetaexpress}
    w_{n} = \frac{\sigma(z_0+(n-1)\kappa)\sigma(z_0+(n+2)\kappa)}{\sigma(\kappa)^4\sigma(z_0+n\kappa)\sigma(z_0+(n+1)\kappa)}
= C\left(\zeta(z_0+(n+1)\kappa)-\zeta(z_0+n\kappa)+\Tilde{c}\right),
\end{equation}
for $C=\frac{\sigma(2\kappa)}{\sigma(\kappa)^4}$ and $\Tilde{c}=\zeta(\kappa)-\zeta(2\kappa)$, 
where the second equality (used in 
\cite{hone2023family} in connection with solutions of the Volterra lattice) follows from 
 addition formulae for Weierstrass functions. We can compare this formula directly with the shadow solution %Notably this has a similar form to the 
$y^{(iv)}_n$. 
\begin{lemma}
    
Up to subtracting multiples of $y^{(i)}_n, y^{(ii)}_n, y^{(iii)}_n$ and overall scale, 
a fourth linearly independent shadow sequence % 
for Somos-5 is $\bar{y}^{(iv)}_n$   given by 
    \begin{equation} \label{Shadow4Solution}
        \bar{y}^{(iv)}_n = x_n\sum^{n-1}_{j=0}w_{j}= C\Big( y^{(iv)}_n +\tilde{c}y^{(iii)}_n -\ze (z_0^{(0)})\big(y^{(i)}_n + y^{(ii)}_n\big)  ) \Big) 
    \end{equation}
    for $w_n= \frac{x_nx_{n+3}}{x_{n+1}x_{n+2}}$ which satisfies
    \begin{equation} \label{wnRecurrence}
    w_{n+1}w_{n-1} = \alpha^{(0)} + \frac{\beta^{(0)}}{w_n},
\end{equation}.
\end{lemma}
\begin{prf}
Using the formula (\ref{zetaexpress}) but with parameters $z_0\to z_0^{(0)}$, $\ka\to \ka^{(0)}$, and so forth, 
we have the telescopic sum 
$$ \begin{array}{rcl}
         \sum^{n-1}_{j=0}w_{j} &= & C\big(\zeta(z_0^{(0)}+n\kappa^{(0)})-\zeta(z_0^{(0)}+(n-1)\kappa^{(0)})+\Tilde{c} + \dots + \zeta(z_0^{(0)}+\kappa^{(0)})-\zeta(z_0^{(0)})+\Tilde{c} \big)\\
         &= & C\big(\zeta(z_0^{(0)}+n\kappa^{(0)}) + n\Tilde{c} -\zeta(z_0^{(0)})\big), 
    \end{array}
$$
and upon multiplying by $x_n$ and comparing with the result of Lemma \ref{shadlem}, we obtain \eqref{Shadow4Solution}. 
\end{prf}
Upon computing another telescopic sum, namely  $x_n\bar{y}^{(iv)}_{n+1}-x_{n+1}\bar{y}^{(iv)}_n$, we find a first order inhomogeneous linear relation for 
this alternative  fourth shadow solution. 
\begin{corollary} The terms of the shadow Somos-5 sequence $(\bar{y}^{(iv)}_n)$ satisfy the relation 
\beq\label{yivreln}
        x_n\bar{y}^{(iv)}_{n+1}=x_{n+1}\bar{y}^{(iv)}_n+x_{n-1}x_{n+2}.
\eeq 
\end{corollary} 
To look into the final sequence $y_n^{(v)}$ we return to the conserved quantity $J$ as defined by the relation \eqref{BaseJ}. 
We noted previously that extending to dual numbers changes this to a dual quantity given by $J=J^{(0)}+J^{(1)}\ve$ where $J^{(0)}$ fulfills the same role as the original conserved quantity and $J^{(1)}$ relates to the $y_n$ sequence and is given by \eqref{DualJDefinition}. We can rewrite this equation in terms of a linear operator acting on $y_n$ by defining
\begin{align}
    L_n &= \sum^4_{j=0}C^{(j)}_nx_{n+j}^{-1}\sh^j, \\
    F_n &= D_n - J^{(1)}x_nx_{n+1}x_{n+2}x_{n+3}x_{n+4},
\end{align}
where $\sh$ is the shift operator that sends $n \rightarrow n+1$. Then \eqref{DualJDefinition} can be rewritten as
\begin{equation} \label{LnFull}
    L_n(y_n)=F_n,
\end{equation}
which, for fixed $J^{(1)}$,  is  a linear inhomogeneous difference equation of order 4 for $y_n$,  reducing the order by 1 from the original recurrence \eqref{oddDualSomos5} 
for $y_n$. \par
The 4th order homogeneous equation that arises when $\alpha^{(1)}=\beta^{(1)}=0$ and $J^{(1)}=0$, namely 
\begin{equation} \label{LnHomogeneous}
    L_n(y_n)=0, 
\end{equation}
is the linearization of the equation defining $J^{(0)}$, whose solutions form a vector space of dimension 4, spanned by the 4 linearly independent shadow solutions 
%obtained by varying the solution with respect to the remaining 4 quantities, ie 
$y_n^{(i)},y_n^{(ii)},y_n^{(iii)},y_n^{(iv)}$. A fifth linearly independent shadow solution for Somos-5, such as $y_n^{(v)}$, then arises from an inhomogenous solution  
of (\ref{LnFull}), by keeping $\alpha^{(1)}=\beta^{(1)}=0$, but taking a non-zero value $J^{(1)}\neq0$. 
%We note that this is because the solution is defined by the invariant $J$ due to it's connection with the elliptic curve. Trying to do the same with the invariant $I$ would require a separate solution in terms of this quantity and 4 other parameters that could not take advantage of this connection. As such these solutions $y_n^{(i)},y_n^{(ii)},y_n^{(iii)},y_n^{(iv)}$ will not have any special value of the invariants $I^{(0)}$ or $I^{(1)}$. \par
The other dual invariant $I$ has even/odd components  $I^{(0)}$ and $I^{(1)}$ which are functionally independent of the components of $J$, so the values taken by $I$ depend 
on the orbits of dual Somos-5; but here we will not  pursue the analogous linear equation obtained from  $I^{(1)}$ any further.

Examining the other sequences $y_n^{(i)},y_n^{(ii)},y_n^{(iii)},y_n^{(iv)}$ as solutions to this homogeneous equation, we can note some identities for the coefficients 
of the operator $L_n$. For $y_n^{(i)}$, $L_n(y_n^{(i)})=0$ gives
\begin{align}
    0=\sum^4_{j=0}C^{(j)}_nx_{n+j}^{-1}y_{n+j} = \begin{cases}
     C_n^{(1)}+C_n^{(3)} \quad&\text{for} \,\, n \,\, \text{odd},\\
     C_n^{(0)}+C_n^{(2)}+C_n^{(4)} \quad &\text{for} \,\, n \,\, \text{even},
    \end{cases} 
\end{align}
The corresponding equations obtained from $L_n(y_n^{(ii)})=0$ are the same but with $n$ odd/even switched, so hence we see 
we must have $ C_n^{(0)}+C_n^{(2)}+C_n^{(4)}=0$ and $C_n^{(1)}+C_n^{(3)}=0$ for all $n$; and from the original definitions of the coefficients $C_n^{(j)}$ in 
\eqref{Cdefs}, we can see 
that  these are satisfied identically. Adding these two shadow solutions, to get the equation  $L_n(x_n)=0$,   
corresponding to the identity  $\sum^4_{j=0}C_n^{(j)}=0$.  Examining $L_n(y_n^{(iii)})=0$ we get the identity 
\begin{equation}
    0= \sum^4_{j=0}C_n^{(j)}(n+j)) =  
     n\sum^4_{j=0}C_n^{(j)}+C_n^{(1)}+2C_n^{(2)}+3C_n^{(3)}+4C_n^{(4)} , 
\end{equation}
and after making use of  $\sum^4_{j=0}C_n^{(j)}=0$ and the other relations, we find that the other three coefficients can be written in terms of $C_n^{(0)}$ and $C_n^{(1)}$, thus: 
\begin{align} \label{CnRelations}
%    C_n^{(0)}, && C_n^{(1)}, &&, 
C_n^{(2)}=-2C_n^{(0)}-C_n^{(1)}, && C_n^{(3)}=-C_n^{(1)}, && C_n^{(4)}=C_n^{(0)}+C_n^{(1)}.
\end{align} \par

The relations \eqref{CnRelations} allow us to rewrite the homogeneous equation \eqref{LnHomogeneous} in a much simpler way, which makes 
the first three independent shadow solutions even more obvious. 
Indeed, if we define  
$$y_n=x_nY_n, $$ 
then  the homogeneous equation becomes 
\begin{equation}
    L_n(y_n)\equiv x_nx_{n+1}x_{n+2}x_{n+3}x_{n+4}\Tilde{L}_n(Y_n)=0,
\end{equation}
%we can write, 
where a short calculation with the above relation yields the 4th order difference operator 
\begin{equation}
    \Tilde{L}_n=\left[ \Tilde{C}_n^{(0)}(\sh+1)+\Tilde{C}_n^{(1)}\sh\right](\sh+1)(\sh-1)^2, 
\end{equation}
where 
\begin{align}
    \Tilde{C}_n^{(0)}&=\alpha^{(0)}\frac{x_{n+1}x_{n+2}}{x_{n}x_{n+3}}+\beta^{(0)}\frac{x_{n+2}^2}{x_nx_{n+4}}-\frac{x_nx_{n+3}}{x_{n+1}x_{n+2}}, \\
    \Tilde{C}_n^{(1)}&=\alpha^{(0)}\left(\frac{x_{n+2}x_{n+3}}{x_{n+1}x_{x+4}}-\frac{x_{n+1}x_{n+2}}{x_nx_{n+3}} \right)+\frac{x_nx_{n+3}}{x_{n+1}x_{n+2}}-\frac{x_{n+1}x_{n+4}}{x_{n+2}x_{n+3}}.
\end{align}
Due to the factorized cubic part  of $\Tilde{L}_n$ with constant coefficients, namely $(\sh+1)(\sh-1)^2$, it is clear that the kernel of this operator has a 3-dimensional subspace 
spanned by $Y_n=1$, $Y_n =n$ and $Y_n=(-1)^n$, and multiplying by $x_n$ we immediately find the subspace of $\ker  L_n$ 
spanned by   
$y_n^{(i)},y_n^{(ii)},y_n^{(iii)}$.

We now present an algebraic formula for the general solution of the 5th order recurrence \eqref{oddDualSomos5} for $y_n$, which is based on 
applying  the method of  variation of parameters in the discrete setting (see \cite{elaydi}) to the 4th order equation \eqref{LnFull}. %This is based on 
The main observation to make initially is that every solution of   \eqref{oddDualSomos5} is also a solution of  \eqref{LnFull}, for some 
value of the first integral $J^{(1)}$. Hence, to solve the original problem, it is 
sufficient to find the general solution of the 4th order equation with an arbitrary parameter $J^{(1)}$. 

For  variation of parameters, we start with any 4 linear independent solutions of the homogeneous equation  \eqref{LnHomogeneous}, $y_n^{(i)},y_n^{(ii)},y_n^{(iii)},y_n^{(iv)}$ say, 
and assume that the solution of the full inhomogeneous problem    \eqref{LnFull} 
takes the form 
\begin{equation}
    y_n = \sum_j f_n^{(j)}y_n^{(j)},
\end{equation}
where $f_n^{(j)}$ are some coefficient functions, as yet undetermined, and  the sum runs over $j=i,ii,iii,iv$. 
Before we substitute this into the equation, we first set the constraints
\begin{equation}\label{kconstr}
    \sum_j(f_{n+1}^{(j)}-f_n^{(j)})y_{n+k}^{(j)}=0 \qquad \mathrm{for} \quad  k=1,2,3, 
\end{equation}
which also implies that 
$$y_{n+k}=\sum_j f_n^{(j)}y_{n+k}^{(j)}, \qquad k=0,1,2,3.$$
 Then substituting this form of $y_n$ into \eqref{LnFull} and applying the constraints (\ref{kconstr}) gives
$$ 
\begin{array}{rcl}
    L_n(y_n)&=& C_n^{(4)}x_{n+4}^{-1}\sum_j (f_{n+1}^{(j)}-f_n^{(j)})y_{n+4}^{(j)} + \sum_j f_n^{j}L_n(y_n^{(j)}) \\
    &=&  C_n^{(4)}x_{n+4}^{-1}\sum_j (f_{n+1}^{(j)}-f_n^{(j)})y_{n+4}^{(j)},
\end{array}
$$ 
as  the shadow sequences $y_n^{(j)}\in\ker L_n$ for $j=i,ii,iii,iv$. Hence, from  \eqref{LnFull}, we have 
$$ 
C_n^{(4)}x_{n+4}^{-1}\sum_j (f_{n+1}^{(j)}-f_n^{(j)})y_{n+4}^{(j)} = F_n, 
$$ 
which, together with the 3 constraints (\ref{kconstr}), gives a system of 4 simultaneous equations for the differences 
$$\delta^{(j)}_n:=f_{n+1}^{(j)}-f_n^{(j)},$$ 
and solving this linear system then determines the functions $f_n^{(j)}$, up to a set of  arbitrary constants $f_0^{(j)}$, corresponding to 
the choice of initial conditions.
\begin{theorem}
    A general solution of \eqref{oddDualSomos5} can be given in the form
    \begin{equation}
    y_n = \sum_j f_n^{(j)}y_n^{(j)},
\end{equation}
where the sum runs over $j=i,ii,iii,iv$ for the 4 shadow sequences $y_n^{(i)},y_n^{(ii)},y_n^{(iii)},y_n^{(iv)}$. The coefficients $f_n^{(j)}$ are given by 
\begin{equation}
    f_n^{(j)}=f_0^{(j)}+\sum^{n-1}_{k=0}\delta^{(j)}_k
\end{equation}
for arbitrary constants %initial 
$f_0^{(j)}$, and 
\begin{equation}
    \begin{pmatrix}
\delta^{(i)}_n \\
\delta^{(ii)}_n \\
\delta^{(iii)}_n \\
\delta^{(iv)}_n
\end{pmatrix} =
\frac{x_{n+4}F_n}{C_n^{(4)}}\begin{vmatrix}
y^{(i)}_{n+1} & y^{(ii)}_{n+1} & y^{(iii)}_{n+1} & y^{(iv)}_{n+1} \\
y^{(i)}_{n+2} & y^{(ii)}_{n+2} & y^{(iii)}_{n+2} & y^{(iv)}_{n+2} \\
y^{(i)}_{n+3} & y^{(ii)}_{n+3} & y^{(iii)}_{n+3} & y^{(iv)}_{n+3} \\
y^{(i)}_{n+4} & y^{(ii)}_{n+4} & y^{(iii)}_{n+4} & y^{(iv)}_{n+4} 
\end{vmatrix}^{-1}
\begin{pmatrix}
d(ii,iii,iv) \\
d(i,iii,iv) \\
d(i,ii,iv) \\
d(i,ii,iii) 
\end{pmatrix},
\end{equation}
where 
\begin{equation}
    d(a,b,c) = \begin{vmatrix}
        y_{n+1}^{(a)} & y_{n+1}^{(b)} & y_{n+1}^{(c)} \\
        y_{n+2}^{(a)} & y_{n+2}^{(b)} & y_{n+2}^{(c)} \\
        y_{n+3}^{(a)} & y_{n+3}^{(b)} & y_{n+3}^{(c)} 
    \end{vmatrix}.
\end{equation}
\end{theorem}
\begin{prf}
 The variation of parameters method reduces the problem to solving the linear system 
$$ 
\left( 
\begin{array}{cccc}
y^{(i)}_{n+1} & y^{(ii)}_{n+1} & y^{(iii)}_{n+1} & y^{(iv)}_{n+1} \\
y^{(i)}_{n+2} & y^{(ii)}_{n+2} & y^{(iii)}_{n+2} & y^{(iv)}_{n+2} \\
y^{(i)}_{n+3} & y^{(ii)}_{n+3} & y^{(iii)}_{n+3} & y^{(iv)}_{n+3} \\
y^{(i)}_{n+4} & y^{(ii)}_{n+4} & y^{(iii)}_{n+4} & y^{(iv)}_{n+4} 
\end{array}
\right) 
 \begin{pmatrix}
\delta^{(i)}_n \\
\delta^{(ii)}_n \\
\delta^{(iii)}_n \\
\delta^{(iv)}_n
\end{pmatrix} 
=  \begin{pmatrix} 
0 \\ 0 \\ 0 \\
(C_n^{(4)})^{-1}x_{n+4}F_n
\end{pmatrix}
. 
$$ 
The four independent shadow sequences $y_n^{(j)}\in\ker L_n$ for $j=i,ii,iii,iv$ can be chosen arbitrarily.  
%and the final linearly independent part is introduced by the choice of the conserved quantity $J^{(1)}$ that appears in $F_n$. 
\end{prf}
We end this section by presenting some explicit examples of shadow sequences.
\begin{Example}\label{ShadowsExplicitExample}
Taking the sequence $x_n$ to be the original well-known Somos-5 sequence \eqref{BaseSomos5Sequence}, i.e. with $\alpha=\beta=x_0=x_1=x_2=x_3=x_4=1$, we 
immediately find the first three shadow sequences in terms of these $x_n$ from \eqref{Shadow1}, \eqref{Shadow2} and \eqref{Shadow3}. For the fourth shadow sequence $y^{(iv)}_n$, 
we can take $\bar{y}^{(iv)}_n$ given by 
%can be found from 
\eqref{Shadow4Solution} in terms of $x_n$ and the quantities $w_n$, which can either be 
calculated directly   from the $x_n$ using \eqref{oldVolterraSolution}, or found recursively using \eqref{wnRecurrence} with $w_1=w_2=1$. The $n=0$ term is fixed to be  
 $\bar{y}^{(iv)}_0=0$ (an empty sum) which is consistent with the relation \eqref{yivreln}. For $y_n^{(v)}$ we must solve \eqref{LnFull} for $\alpha^{(1)}=\beta^{(1)}=0$ and $J^{(1)}\neq0$. For simplicity we start with $y_n^{(v)}=0$ for $n=0,1,2,3$ and set $J^{(1)}={-1}$ to cancel the $-$ sign in $F_n$, which gives $y_4^{(v)}=1$, and 
the rest of the sequence is found by continuing with the definition of $J^{(1)}$ or by iterating the base recurrence \eqref{linearizedOddSomos5}. 
Numerical values up to $n=10$ are given in Table \ref{ExamplesTable}.
\begin{table}[ht!]
\centering
 \begin{tabular}{||c c c c c c c c c c c c||} 
 \hline
 n & 0 & 1 & 2 & 3 & 4 & 5 & 6 & 7 & 8 & 9 & 10 \\ [0.5ex] 
 \hline\hline
 $x_n$ & 1 & 1 & 1 & 1 & 1 & 2 & 3 & 5 & 11 & 37 & 83\\ 
 $y_n^{(i)}$ & 1 & 0 & 1 & 0 & 1 & 0 & 3 & 0 & 11 & 0 & 83\\
 $y_n^{(ii)}$ & 0 & 1 & 0 & 1 & 0 & 2 & 0 & 5 & 0 & 37 & 0 \\
 $y_n^{(iii)}$ & 0 & 1 & 2 & 3 & 4 & 10 & 18 & 35 & 88 & 333 & 830 \\
 $y_n^{(iv)}$ & 0 & 2 & 3 & 4 & 6 & 15 & 25 & 49 & 130 & 475 & 1147 \\
 $y_n^{(v)}$ & 0 & 0 & 0 & 0 & 1 & 1 & 2 & 5 & 17 & 23 & 118 \\ [1ex] 
 \hline
 \end{tabular}
 \caption{The original Somos-5 sequence \eqref{BaseSomos5Sequence} 
%for $x_0=x_1=x_2=x_3=x_4=1=\alpha=\beta$ 
and 5 linearly independent Shadow sequences associated with it.}
 \label{ExamplesTable}
\end{table} \par
The 5 sequences are linearly independent, as can be seen clearly for these values by considering the initial conditions as $y_0,y_1,y_2,y_3,y_4$ and the 
corresponding vectors $\textbf{y}^{(j)}=(y^{(j)}_0,y^{(j)}_1,y^{(j)}_2,y^{(j)}_3,y^{(j)}_4)$ for $j=i,ii,iii,iv,v$. Then we can explicitly expand the standard basis for $\C^5$ 
in terms of these vectors as 
\begin{align*}
    \textbf{e}_1&=\textbf{y}^{(i)}+\textbf{y}^{(v)}-\textbf{y}^{(iv)}+\textbf{y}^{(ii)}+\textbf{y}^{(iii)}, \\
    \textbf{e}_2&=\textbf{y}^{(iv)}-\frac{3}{2}\textbf{y}^{(iii)}+\frac{1}{2}\textbf{y}^{(ii)}, \\
    \textbf{e}_3&=\textbf{y}^{(iv)}-2\textbf{y}^{(v)}-\textbf{y}^{(ii)}-\textbf{y}^{(iii)}, \\
    \textbf{e}_4&=\frac{1}{2}\textbf{y}^{(ii)}-\textbf{y}^{(iv)}+\frac{3}{2}\textbf{y}^{(iii)}, \\
    \textbf{e}_5&= \textbf{y}^{(v)},\\
\end{align*} 
which shows that 
$\{\textbf{y}^{(i)},\textbf{y}^{(ii)},\textbf{y}^{(iii)},\textbf{y}^{(iv)},\textbf{y}^{(v)}\}$ span the space of initial conditions 
(although not as a $\Z$-module). 
% and we can explicitly find the unit vectors as (BUT not a \Z-basis).
\end{Example}

\section{Hankel determinant formulae}

Using a combinatorial approach and determinant identities,  Hankel determinant formulae for Somos-5 sequences were derived in \cite{chang}, while in recent work on discrete integrable systems related to the Volterra lattice, a different set of Hankel determinant formulae were found by one of us  via the connection with Stieltjes continued fractions 
 \cite{hone2023family}. In the notation used in the latter work, the sequence of Hankel determinants is specified by  
\beq\label{hdets}  \Delta_{2k-1} = \det(s_{i+j-1})_{i,j=1,2,\dots,k}, \qquad  \Delta_{2k} = \det(s_{i+j})_{i,j=1,2,\dots,k}\eeq 
for $k\geq 1$, with $\Delta_{-2}=\Delta_{-1}=\Delta_0=1$, where the entries $s_j$ (the moments) are defined by the recursion relation  
\begin{equation}\label{srec} s_j = \left(-2w_1- \frac{c_1}{2}\right)s_{j-1}+ \sum^{j-1}_{i=1}s_is_{j-i} + \left(\frac{1}{4w_1}\left(\frac{c_1^2}{4}-c_2 \right)-w_2 \right)\sum^{j-2}_{i=1}s_is_{j-i-1}\end{equation}
for $j\geq 3$, with initial values $s_1 = w_1, s_2 = w_1w_2$. Here, as usual,  the $w_n$ are given by the recurrence \eqref{OldVolterraMap}, which is the QRT map associated with  the Somos-5 relation \eqref{origs5} via the substitution 
\beq\label{wnewsub}w_n = \frac{x_nx_{n+3}}{x_{n+1}x_{n+2}}, \eeq  
(and the reader should note that in this section we have made an overall shift of index compared with \eqref{oldVolterraSolution}, in order to be consistent with the conventions used in  \cite{hone2023family}). The remaining constants $c_1, c_2$ can be fixed from a related elliptic curve, or by a variety of relations to the other parameters. They can be written
$$ 
\begin{array}{rcl}
    c_1 &=& -2\left(\frac{\alpha}{w_0} + w_1 + w_0 + w_{-1} \right) =-2J  \\
    c_2 &= &4\alpha+\frac{c_1^2}{4}= 4\alpha + J^2 = 12\Tilde{\mu}\tilde{\lambda},
\end{array}
$$ 
where the substitution \eqref{wnewsub} has been used to obtain the extra equalities in each line above,  rewriting $c_1,c_2$ in terms of other constants used earlier, namely the first integral %%conserved quantity 
$J$ defined in \eqref{BaseJ}, and 
$\Tilde{\mu}$, 
$\tilde{\lambda}$ as in \eqref{SolutionConstants}. The $w_n$ can then be given in terms of these determinants by
$$ w_n = \frac{\Delta_{n-3}\Delta_n}{\Delta_{n-1}\Delta_{n-2}} = \frac{x_nx_{n+3}}{x_{n+1}x_{n+2}},$$ 
so the $\Delta_n$ generate a Somos-5 sequence with $x_1=x_2=x_3=1, x_4=s_1, x_5=s_2$, if we identify   
$$x_n=\Delta_{n-3}.$$

\begin{Example}\label{cpxs5h} 
For the original Somos-5 sequence we have $x_0=x_1=x_2=x_3=x_4=\alpha=\beta=1$ and $J=5$, thus $s_1 = w_1 = 1$ and $s_2= w_1w_2=2$. From the above the entries of 
the Hankel matrices are found recursively using %can be given by 
\begin{equation}
    s_j = 3s_{j-1}+ \sum^{j-1}_{i=1}s_is_{j-i} -3\sum^{j-2}_{i=1}s_is_{j-i-1}, 
\end{equation}
which recovers Example 3.9 in \cite{hone2023family}, and the Hankel determinants $\Delta_n = x_{n+3}$ can be seen to reproduce 
the original Somos-5 sequence \eqref{BaseSomos5Sequence}. 
\end{Example} 
In  \cite{hone2023family}, integrable maps and Hankel determinant solutions are constructed from continued fraction expansions on hyperelliptic curves of arbitrary genus $g$, but for what follows it will be convenient to paraphrase the main results about Hankel determinants in the case $g=1$, which correspond to Somos-5 sequences in the following way.   
%Additionally for the regular Somos-5 case, we can rearrange the above and state a form in which $s_1,s_2$ and the coefficients in the relation are taken as initial conditions.
\begin{proposition}\label{hankg1} Suppose that a sequence $(\Delta_n)_{n\geq -2}$ is specified by 
$\Delta_{-2}=\Delta_{-1}=\Delta_0=1$, and  Hankel determinants $\Delta_{2k-1}$, $\Delta_{2k}$ given by \eqref{hdets} for $k\geq 1$, 
%\begin{align} \Delta_{2k-1} = \det(s_{i+j-1})_{i,j=1,2,\dots,k} && \Delta_{2k} = \det(s_{i+j})_{i,j=1,2,\dots,k}, \qquad k\geq 1 ,\end{align}
with entries given by the moment sequence $(s_j)$ obtained from the recursion \eqref{srec} for $j\geq 3$, 
%\begin{equation}   s_j = k_1s_{j-1}+ \sum^{j-1}_{i=1}s_is_{j-i} + k_2\sum^{j-2}_{i=1}s_is_{j-i-1}, \qquad j\geq 3, \end{equation}
for a fixed choice of constants $k_1,k_2$ and  initial values $s_1, s_2$.  Then $\Delta_n$ is a solution of the Somos-5 recurrence 
%\begin{equation}
$$
    \Delta_{n+3}\Delta_{n-2} = \alpha\Delta_{n+2}\Delta_{n-1} + \beta\Delta_{n+1}\Delta_{n}, \qquad n\geq 0, 
$$ %\end{equation}
where the coefficients $\alpha$ and $\beta$ are given by
\begin{align}
    \alpha = -k_2s_1-s_2, && \beta = k_2s_2+k_1s_2 + k_2s_1^2+2s_1s_2,
\end{align}
and the resulting value of the first integral $J$ is % given by
\begin{equation}
    J = k_1+2s_1.
\end{equation}
\end{proposition} 
Note that fixing initial values $\Delta_{-2}=\Delta_{-1}=\Delta_0=1$ means that there are just 4 free parameters $s_1,s_2,k_1,k_2$, compared with the  7 parameters required 
to specify the general initial value problem for Somos-5. 
% appearing in the base recurrence for a Somos-5 sequence. 
However, other initial values for the sequences can be obtained, and the missing degrees of freedom can be restored, by applying the 3-parameter group of 
scaling symmetries
\begin{align} \label{ScalingSymmetries}
    x_{2k-1} \rightarrow A_{-}x_{2k-1},\qquad \qquad & x_{2k}\rightarrow A_+ x_{2k}, & x_n \rightarrow B^{n}x_n
\end{align}
for arbitrary non-zero constants $A_-, A_+,B\in\C^*$. % for odd/even $n$ respectively. \par

Hankel determinant formulae for negative  values of the index $n$ %sequence 
are also presented in \cite{hone2023family}, given by a similar formula and recursion for the entries. The scaling symmetries above 
can be applied in order to ensure the positive and negative index  formulae line up to form a full sequence for all $n\in\Z$. \par
As with the explicit form of the Somos-5 solution previously, the above Hankel determinant formula extends directly to the dual numbers, because  $\mathbb{D}$ is 
a commutative ring, simply by taking all variables and constants to be in $\D$. %dual terms.
\begin{proposition}
Suppose that constants $k_1,k_2\in \mathbb{D}$ and $s_1, s_2\in \mathbb{D^*}$ are given, and $\Delta_{-2}=\Delta_{-1}=\Delta_0=1$. 
Then the sequence $(\Delta_n)$ defined from the Hankel determinants  
\begin{align}
     \Delta_{2k-1} = \det(s_{i+j-1})_{i,j=1,2,\dots,k} && \Delta_{2k} = \det(s_{i+j})_{i,j=1,2,\dots,k}, \qquad k\geq 1, 
\end{align}
with entries given by the sequence of moments $s_j = s_j^{(0)}+s_j^{(1)}\ve \in \mathbb{D}$ generated recursively from 
\begin{equation}
    s_j = k_1s_{j-1}+ \sum^{j-1}_{i=1}s_is_{j-i} + k_2\sum^{j-2}_{i=1}s_is_{j-i-1}, \qquad j\geq 3, 
\end{equation}
is  a solution of the dual Somos-5 recurrence 
\begin{equation} \label{dualdelta} 
 \Delta_{n+3}\Delta_{n-2} = \alpha\Delta_{n+2}\Delta_{n-1} + \beta\Delta_{n+1}\Delta_{n}, \qquad n\geq 0,  
\end{equation}
where the coefficients $\alpha = \alpha^{(0)}+\alpha^{(1)}\ve$ and $\beta = \beta^{(0)}+\beta^{(1)}\ve$ are given by
\begin{align} \label{alphabeta0}
    \alpha^{(0)} = -k_2^{(0)}s_1^{(0)}-s_2^{(0)}, && \beta^{(0)} = k_2^{(0)}s_2^{(0)}+k_1^{(0)}s_2^{(0)} + k_2^{(0)}\left(s_1^{(0)}\right)^2+2s_1^{(0)}s_2^{(0)},
\end{align}
and
\begin{align}
    \alpha^{(1)}&=-k_2^{(0)}s_1^{(1)}-k_2^{(1)}s_1^{(0)}-s_2^{(1)},\label{alpha1} \\
    \beta^{(1)} &= k_2^{(0)}s_2^{(1)} +k_2^{(1)}s_2^{(0)}+k_1^{(1)}s_2^{(0)}+k_1^{(0)}s_2^{(1)}+k_2^{(1)}\left(s_1^{(0)}\right)^2 + 2k_2^{(0)}s_1^{(1)}s_1^{(0)}+2s_1^{(1)}s_2^{(0)} + 2s_1^{(0)}s_2^{(1)}. \label{beta1}
\end{align}
and the components of the resulting %Somos-5 sequence has 
dual first integral $J = J^{(0)}+J^{(1)}\ve$  
take the values %given by
\begin{equation} \label{Jconstant}
    J^{(i)} = k_1^{(i)}+2s_1^{(i)} \quad \text{for } i=0,1.
\end{equation}
\end{proposition}
\par
The constants $k_1,k_2$ can also be given explicitly in terms of an initial set of  values of $x_n, y_n$, % as well as being found through the relations in the definition, 
via the formulae 
\begin{align}
    k_1^{(0)} = J^{(0)}-2\frac{x_1x_4}{x_2x_3}, && k_2^{(0)} = \frac{x_1x_4}{x_2x_3}+\frac{x_0x_3}{x_1x_2}-J^{(0)},
\end{align}
and
\begin{align}
    k_1^{(1)} &= J^{(1)}-2\frac{y_1x_4}{x_2x_3}-2\frac{x_1y_4}{x_2x_3}+2\frac{x_1y_2x_4}{x_2^2x_3}+2\frac{x_1y_3x_4}{x_2x_3^2},\\
    k_2^{(1)} &= \frac{y_1x_4}{x_2x_3}+\frac{x_1y_4}{x_2x_3}-\frac{x_1y_2x_4}{x_2^2x_3}-\frac{x_1y_3x_4}{x_2x_3^2}+\frac{y_0x_3}{x_1x_2}+\frac{x_0y_3}{x_1x_2}-\frac{x_0y_1x_3}{x_1^2x_2}-\frac{x_0y_2x_3}{x_1x_2^2}-J^{(1)},
\end{align}
which can be obtained via the definitions of $k_1,k_2$ in  terms of the $w_n$ and $J$ from \cite{hone2023family}, 
and using \eqref{dualReciprocal} to expand into even/odd components after changing all variables into dual numbers. 
Assuming $x_1=x_2=x_3=1$ and $y_1=y_2=y_3=0$ to match up with the initial terms of the sequence $(\Delta_n)$, with 
%as in the unscaled definition of the Hankel determinant to get 
$\Delta_{-2}=\Delta_{-1}=\Delta_{0}= 1$, this  simplifies considerably,  to give 
\begin{align} \label{kzerocoefficients}
    k_1^{(0)} = J^{(0)}-2x_4, && k_2^{(0)} = x_4+x_0-J^{(0)},
\end{align}
and
\begin{align} \label{konecoefficients}
    k_1^{(1)} = J^{(1)}-2y_4 && k_2^{(1)} =y_4+y_0-J^{(1)},
\end{align} \par
The Hankel determinant formulae can now be demonstrated to recover some of the dual  Somos-5  sequences that were found previously
in the literature. 
\begin{Example}
    To obtain the Shadow sequence $y^{(v)}_n$ as in Example \ref{ShadowsExplicitExample}, we have $\alpha^{(0)}=\beta^{(0)}=x_0=x_1=x_2=x_3=x_4=1$ and $\alpha^{(1)}=\beta^{(1)}=y_0=y_1=y_2=y_3=0$ with $J^{(1)}=-1$. Hence $\Delta_{-2}=\Delta_{-1}=\Delta_0=1$ from $\Delta_n=x_{n+3}+y_{n+3}\ve$ as required. Then $J^{(0)}=5$ can be found from \eqref{BaseJ} using the initial values of $x_n$. 
    Hence $s_1^{(0)}=x_4=1$ via $w_1 = \frac{X_1X_4}{X_2X_3}$ and the reciprocal formula \eqref{dualReciprocal}. 
Similarly, $s_2^{(0)}=x_5=2$ which can be found using the recurrence for the $x_n$. Then from the $y_n$ values, we find $s_1^{(1)}=1, s_2^{(1)}=1$. 
    We can make use of the relations 
for the components of $\alpha$ and $\beta$ to get the coefficients $k_1,k_2$. In fact, without using 
 \eqref{alphabeta0}, we already know  from Example \ref{cpxs5h} that we have $k_1^{(0)}=3$ and $k^{(0)}_2=-3$, because the even parts 
must agree with the parameters for the ordinary Somos-5 over $\C$. 
From the $n=1$ case of \eqref{Jconstant} we find $k_1^{(1)}=-3$ and via \eqref{alpha1} we get $k_2^{(1)}=2$. Hence
    \begin{equation}
    s_j = (3-3\ve)s_{j-1}+ \sum^{j-1}_{i=1}s_is_{j-i} - (3-2\ve)\sum^{j-2}_{i=1}s_is_{j-i-1}, \qquad j\geq 3, 
\end{equation}
with  initial moments  $s_1= 1+\varepsilon, s_2=2+\varepsilon$. 
From this we can find the first few dual moments:  $s_3=7-\varepsilon, s_4=27-18\varepsilon,s_5=109-119\varepsilon$, and  verify that the determinants recover the sequence. We have $\Delta_1=s_1=1+\varepsilon$, $\Delta_2=2+\varepsilon$ and
\begin{align}
    \Delta_3 &= \begin{vmatrix}
1+\varepsilon & 2+\varepsilon \\
2+\varepsilon & 7-\varepsilon
\end{vmatrix} = 3+2\varepsilon,\\
\Delta_4&= \begin{vmatrix}
2+\varepsilon & 7-\varepsilon \\
7-\varepsilon & 27-18\varepsilon
\end{vmatrix} = 5+5\varepsilon,\\
 \Delta_5 &= \begin{vmatrix}
1+\varepsilon & 2+\varepsilon & 7-\varepsilon \\
2+\varepsilon & 7-\varepsilon & 27-18\varepsilon \\
7-\varepsilon & 27-18\varepsilon & 109-119\varepsilon
\end{vmatrix} = 11+17\varepsilon.
\end{align}
So from the even parts we can see the expected sequence  $x_n$: $1,1,1,1,1,2,3,5,11,\dots$, while the odd parts give the shadow sequence $y^{(v)}_n$ as $y_n$: $0,0,0,0,1,1,2,5,17,\dots$. 
\end{Example} \par

We can also demonstrate the use of the scalings \eqref{ScalingSymmetries} to get other dual sequences, when we consider 
an example of one of  Ovsienko and Tabachnikov's specific cases, namely \eqref{OvsienkoForm1}. 

\begin{Example}
    From \eqref{OvsienkoForm1} we have $\alpha = 1 + \alpha^{(1)}\varepsilon$, $\beta=1$. Letting $\alpha^{(1)}=2$ for this example, with the usual initial conditions $x_0=x_1=x_2=x_3=x_4=1$ and $y_0=y_1=y_2=y_3=y_4=0$ so that $\Delta_{-2}=\Delta_{-1}=\Delta_{0}= 1$, we have from \eqref{kzerocoefficients} that $k_1^{(0)}=3, k_2^{(0)}=-3$. Similarly we can find $J^{(1)}=6$ from \eqref{DualJDefinition} and thus, via \eqref{konecoefficients}, $k_1^{(1)}=4, k_2^{(1)}=-4$. Also have $s_1=X_4=1$, and using the full recurrence for the $y_n$, \eqref{oddDualSomos5}, with the above values we find $y_5=2$ so $s_2=X_5=2+2\varepsilon$. Hence the entries can be generated by
    \begin{equation}
         s_j = (3+4\varepsilon)s_{j-1}+ \sum^{j-1}_{i=1}s_is_{j-i} - (3+4\varepsilon)\sum^{j-2}_{i=1}s_is_{j-i-1},
    \end{equation}
    which gives $s_3=7+14\varepsilon, s_4=27+78\varepsilon, s_5=109+402\varepsilon$. Hence
    \begin{align}
    \Delta_3 &= \begin{vmatrix}
1 & 2+2\varepsilon \\
2+2\varepsilon & 7+14\varepsilon
\end{vmatrix} = 3+6\varepsilon,\\
\Delta_4&= \begin{vmatrix}
2+2\varepsilon & 7+14\varepsilon \\
7+14\varepsilon & 27+78\varepsilon
\end{vmatrix} = 5+14\varepsilon,\\
 \Delta_5 &= \begin{vmatrix}
1 & 2+2\varepsilon & 7+14\varepsilon \\
2+2\varepsilon & 7+14\varepsilon & 27+78\varepsilon \\
7+14\varepsilon & 27+78\varepsilon & 109+402\varepsilon
\end{vmatrix} = 11+42\varepsilon.
\end{align}
Hence we can see the expected sequence  $x_n$: $1,1,1,1,1,2,3,5,11,\dots$, and we have the  sequence of odd parts $y_n$: $0,0,0,0,0,2,6,14,42,\dots$ which can be checked via the recurrence \eqref{oddDualSomos5}. 
We can also demonstrate rescaling using \eqref{ScalingSymmetries} to get sequences with initial values other than $\Delta_{-2}=\Delta_{-1}=\Delta_0=1$. For example, to obtain the initial conditions $y_n=n$ for $n=0,\dots,4$ we can use $A_-=A_+=1$ and $B=1+\varepsilon$. Then by scaling the $X_n=x_n+y_n\varepsilon$ for the sequences above to $X_n \rightarrow B^nX_n$, we obtain the same sequence $x_n$: $1,1,1,1,1,2,3,5,11,\dots$ but the alternative odd component sequence $y_n$: $0,1,2,3,4,12,24,49,130,\dots$ which can again be verified by \eqref{oddDualSomos5}. Hence sequences with other initial conditions can still be obtained from the original Somos-5 sequence, by using the Hankel determinant formulae 
together with appropriate scalings.
\end{Example}

\section{Bi-Hamiltonian structure for dual QRT map}

Iterating the recurrence (\ref{OldVolterraMap}) is equivalent to making iterations of the  birational  map 
\beq\label{s5qrt} 
\varphi: \quad \left(\begin{array}{c} w_1 \\ 
w_2 \end{array}\right) \longmapsto 
\left(\begin{array}{c} w_2 \\ 
w_1^{-1}(\al +\be w_2^{-1}) \end{array}\right), 
\eeq 
in the affine  plane with coordinates $(w_1,w_2)$, 
which preserves the log-canonical symplectic form 
$$ 
\om = \frac{\dd w_1\wedge \dd w_2}{w_1w_2},  
$$ 
in the sense that $\varphi^*(\om)=\om$. 
The first integral $J$ for Somos-5 can be rewritten  in terms 
of a pair of  coordinates $(w_1,w_{2})$ in the plane, 
by fixing $n=1$ in 
(\ref{biqJ}), 
to obtain 
$$ 
J= w_1 +w_2 +\al \left(\frac{1}{w_1}+\frac{1}{w_2}\right) + \frac{\be}{w_1w_2}, 
$$  
and the family of level sets of $J$ defines a pencil of biquadratic plane curves. 
The map (\ref{s5qrt}) is a particular example of a symmetric QRT map, and is associated with elliptic solutions 
of the Volterra lattice (see \cite{hone2023family} and references). 

\begin{figure}[h]\label{orbit1}
  \centering 
  \epsfig{file=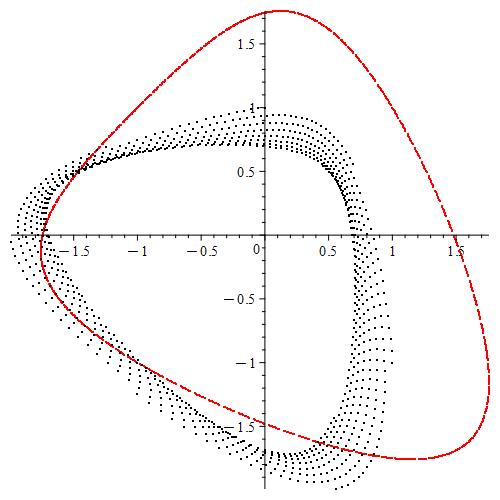, height=3.7in, width=3.7in}
\caption{Plot in $(v_1,v_2)$ plane of 1000 points on two different shadow orbits of the map $\varphi_{dual}$ for 
$\al=\be=1$ and $(w_1,w_2)=(1,1)$ with initial  values $(v_1,v_2)=(-1,1)$ and $(v_1,v_2)=(0,1)$,   
respectively} 
\end{figure} 

As already mentioned above, the substitution (\ref{oldVolterraSolution}) relates solutions of the Somos-5 recurrence to iterates of the map
(\ref{s5qrt}), and we can extend this to the dual numbers by setting 
\beq\label{dualsub}
W_n = \frac{X_{n+2}X_{n-1}}{X_{n+1}X_n}, 
\eeq
where $X_n$ is a solution of (\ref{dualsomos5}). 
Upon expanding into even/odd components we find 
$$ 
W_n =w_n +\ve v_n, 
$$ 
where the even component $w_n$ corresponds to iterates of $\varphi$ as before, but with the replacement 
$\al\to\al^{(0)}$, $\be\to\be^{(0)}$ in (\ref{s5qrt}),  and the odd components are given in terms of $x_n$ and $y_n$ by 
\beq\label{vxy}
v_n = \frac{x_{n-1}y_{n+2}}{x_{n+1}x_n} 
- \frac{x_{n-1}x_{n+2}y_{n+1}}{x_{n+1}^2x_n} 
- \frac{x_{n-1}x_{n+2}y_{n}}{x_{n+1}x_n^2} 
+  \frac{x_{n+2}y_{n-1}}{x_{n+1}x_n}
.
\eeq 
Moreover, the odd components satisfy the recurrence   
\beq\label{vrec}
w_{n-1}w_n v_{n+1} + w_{n-1}w_{n+1}v_n + w_{n}w_{n+1}v_{n-1}  = \al^{(0)} \, v_n +\al^{(1)}\, w_{n}+   \be^{(1)}, 
\eeq 
which comes from the odd part of the  recurrence  for $W_n$,   that is 
\beq\label{dualqrtrec} 
W_{n+1}W_nW_{n-1} = \al \, W_n +\be, \qquad \al,\be\in\D. 
\eeq 
The above recurrence corresponds to the 
dual QRT map
\beq\label{ds5qrt} 
\varphi_{dual}: \quad \left(\begin{array}{c} W_1 \\ 
W_2 \end{array}\right) \longmapsto 
\left(\begin{array}{c} W_2 \\ 
W_1^{-1}(\al  +\be W_2^{-1}) \end{array}\right), \qquad (W_1,W_2)\in\D^*,  
\eeq 
which can be written in components as the 4D map 
\beq\label{dualqrt} 
%\varphi_{dual}: \quad 
\left(\begin{array}{c} w_1 \\ w_2 \\ v_1 \\ v_2 \end{array}\right) \longmapsto 
\left(\begin{array}{c} w_2 \\ w_3 \\ v_2 \\ v_3 \end{array}\right), \qquad \mathrm{where} \quad  
w_3 = w_1^{-1}w_2^{-1}(\al^{(0)} w_2 +\be^{(0)})   , 
\eeq 
and 
$$ 
v_3  = -w_2^{-1}w_3 v_2 -w_1^{-1}w_3 v_1 
+w_1^{-1}w_2^{-1}\big( \al^{(0)} v_2 + \al^{(1)} w_2+ \be^{(1)}\big) . 
$$ 

\begin{remark} 
In the general setting of noncommutative  variables, %in case that  
where $W_1,W_2$ are considered as units in an associative algebra, with parameters $\al,\be$ both set to 1, 
Duzhin and Kontsevich discovered the map 
(\ref{ds5qrt}) in the form 
$$ (W_1,W_2)\mapsto \big(W_1W_2W_1^{-1}, (1+W_2^{-1})W_1^{-1}\big),  
$$  
which they found was a discrete symmetry of the ODE system 
\beq\label{flow}
\dot{W}_1 = W_1W_2-W_1W_2^{-1} - W_2^{-1}, \quad 
 \dot{W}_2 = -W_2W_1+W_2W_1^{-1} + W_1^{-1}. 
\eeq 
(See \cite{nonc} and \cite{wolf} for more details.) 
\end{remark} 

We now derive a Poisson structure that is compatible with the dual QRT map $\varphi_{dual}$, which is  obtained simply by noticing that the dual analogue of the symplectic form $\om$, 
namely 
\beq\label{bigom} 
\Omega =   \frac{\dd W_1\wedge \dd W_2}{W_1W_2},  
\eeq 
is preserved by (\ref{ds5qrt}), so that $\varphi^*_{dual}(\Omega)=\Omega$. Then  we can expand $\Omega$ into its even and odd components, 
$$ 
\Omega= \omega^{(0)} + \ve \,   \omega^{(1)}, 
$$ 
where 
$$ 
 \omega^{(0)}= \om 
$$ 
is the same as the original log-canonical 2-form on the $(w_1,w_2)$ plane, while
$$
 \omega^{(1)} = -\left(\frac{v_1}{w_1}+ \frac{v_2}{w_2}\right) \, \om + \frac{1}{w_1w_2} \, \big( 
\dd v_1\wedge \dd w_2 + \dd w_1\wedge\dd v_2
\big), 
$$ 
and note that each component is separately invariant under the map in the extended 4D phase space: 
$$ 
\varphi^*_{dual}( \omega^{(j)} ) =  \omega^{(j)}, \quad j=0,1.
$$

\begin{figure}[h]\label{orbit2}
  \centering 
  \epsfig{file=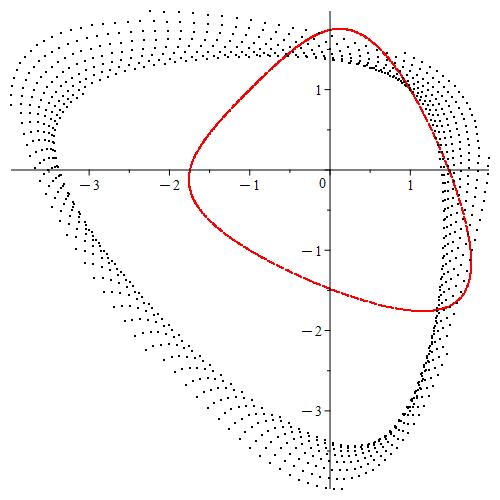, height=3.7in, width=3.7in}
\caption{The same as in Figure 1, %\ref{orbit1}, 
but now comparing the orbits   
with initial  values $(v_1,v_2)=(-1,1)$ and $(v_1,v_2)=(1,1)$,   
respectively} 
\end{figure} 

\begin{theorem} The dual QRT map $\varphi_{dual}$ defined by (\ref{dualqrt}) is bi-Hamiltonian, in the sense that it preserves the pencil  of Poisson brackets $\{\, , \, \}$  
with parameter $\zeta$ defined by 
\beq\label{pbze} 
\begin{array}{ll}   
&\{ w_1,w_2\} = \{w_1,v_1\}=\{w_2,v_2\} = 0 , \\ \quad 
&\{w_2,v_1\}=-\{w_1,v_2\} = w_1w_2, \\ 
&\{v_1,v_2\} = w_1w_2\, \zeta - w_1v_2-w_2v_1. 
\end{array} 
\eeq  
Furthermore, the even and odd components of the first integral $J$ are in involution with respect to this pencil, hence $\varphi_{dual}$ is a Liouville integrable map in 4D. 
\end{theorem} 
\begin{prf}The form (\ref{bigom}) can be written as $\dd \log W_1\wedge \dd\log W_2$, so $\dd\Omega=0$ which implies that the even and odd parts are both 
closed. Hence $\omega^{(0)}$ and $\omega^{(1}$ are both symplectic forms that are preserved by the map $\varphi_{dual}$, as is the linear combination 
$$ 
\om_\ze = \zeta \omega^{(0)}+ \omega^{(1)}. 
$$ 
The 2-form $\om_\ze$ is nondegenerate for any $\ze\in\C$. Hence its inverse defines a nondegenerate Poisson tensor, which corresponds to the pencil of brackets 
 $\{\, , \, \}$  given in (\ref{pbze}), and  $ \varphi_{dual}^*(\om_\ze)=\om_\ze$ implies that this bracket is preserved by the dual QRT map for all $\ze$. In the form 
(\ref{ds5qrt}), it is clear that the map preserves the invariant $J\in\D$ given by the dual analogue of (\ref{biqJ}), that is 
$$ 
J= W_1 +W_2 +\al \left(\frac{1}{W_1}+\frac{1}{W_2}\right) + \frac{\be}{W_1W_2}= J^{(0)}+\ve\, J^{(1)},  
$$ 
where 
\beq\label{jints}
\begin{array}{rcl} 
J^{(0)}& =& w_1 +w_2 +\al^{(0)} \left(\frac{1}{w_1}+\frac{1}{w_2}\right) + \frac{\be^{(0)}}{w_1w_2},  \\
J^{(1)}& = & 
\left(1-\frac{\al^{(0)}}{w_1^2} -\frac{\be^{(0)}}{w_1^2w_2} \right)\, v_1 
+ 
 \left(1-\frac{\al^{(0)}}{w_2^2} -\frac{\be^{(0)}}{w_1w_2^2} \right)\, v_2 
+ \al^{(1)} \left(\frac{1}{w_1}+\frac{1}{w_2}\right) + \frac{\be^{(1)}}{w_1w_2},
\end{array} 
\eeq 
which correspond to rewriting the conserved quantities (\ref{j0exp}) and (\ref{DualJDefinition}) for the dual Somos-5 recurrence in terms of 
the coordinates for the 4D phase space. 
The latter define two independent invariants for the map  in 4D, and $\{J^{(0)},J^{(1)}\}=0$ for all $\ze \in\C$, hence the map 
$\varphi_{dual}$ is integrable in the Liouville sense. 
\end{prf} 
\begin{remark} Because they are in involution, the conserved quantities (\ref{jints}) generate a pair of commuting vector fields from any member of the Poisson pencil. Fixing 
$\ze=0$ to obtain the bracket corresponding to the symplectic form $\om^{(1)}$, the first vector field is $\{ \cdot , J^{(0)}\}|_{\ze=0}$ given by 
$$ 
\begin{array}{rcl} 
w_1'& =&  0, \\   w_2' & =& 0, \\ 
 v_1' & = &  - w_1w_2 + (\al^{(0)} w_1 +\be^{(0)})w_2^{-1},  \\
v_2' & = & w_1w_2 - (\al^{(0)} w_2 +\be^{(0)})w_1^{-1}  
\end{array}
$$
(with the derivative denoted by prime), while the second vector 
field is $\{ \cdot , J^{(1)}\}|_{\ze=0}$ given by 
$$ 
\begin{array}{rcl} 
\dot{w}_1& =&  -w_1w_2  + (\al^{(0)} w_1 +\be^{(0)})w_2^{-1}, \\   
\dot{w}_2 & =& w_1w_2  - (\al^{(0)} w_2 +\be^{(0)})w_1^{-1}, \\ 
 \dot{v}_1& = &  (- w_2 + \al^{(0)} w_2^{-1}) v_1 
-(w_1 + \al^{(0)} w_1 w_2^{-2}+ \be^{(0)}  w_2^{-2}) v_2 
+  (\al^{(1)} w_1 + \be^{(1)})  w_2^{-1} , 
 \\
\dot{v}_2 & = & (w_2 + \al^{(0)} w_2 w_1^{-2}+ \be^{(0)}  w_1^{-2}) v_1 
+  ( w_1 - \al^{(0)} w_1^{-1}) v_2  
-  (\al^{(1)} w_2 + \be^{(1)})  w_1^{-1} . 
\end{array}
$$
In terms of the original dual coordinates, the latter is 
$$ 
\dot{W}_1 = -W_1W_2 +(\al W_1+\be )W_2^{-1}, \quad 
\dot{W}_2 = W_1W_2 -(\al W_2+\be )W_1^{-1}, 
$$ 
which, when $\al=\be=1$,  corresponds to (\ref{flow})
(up to an overall minus sign).  Observe that the first two components  $\dot{w}_1$, $\dot{w}_2$ above are the Hamiltonian vector field produced  
by $J^{(0)}$  with the bracket $\{w_1,w_2\}=w_1w_2$, corresponding to the form $\om$ in the $(w_1,w_2)$ plane, but this form  
becomes degenerate when it is extended to the 4D phase space. 
\end{remark} 

The map $\varphi_{dual}$ provides a geometric interpretation of the shadow Somos-5 sequences. If we take the additional parameters 
$\al^{(1)}=\be^{(1)}=0$, then  (\ref{dualqrt}) is just the original QRT map together with its linearization, and the points $(v_1,v_2)$ 
are the shadow in the plane. We can obtain a sequence of such points by using the formula (\ref{vxy}), where $(x_n)$ is an ordinary Somos-5 sequence 
and $(y_n)$ is any one of the shadow sequences $(y_n^{(j)})$ for $j=i,ii,iii,iv,v$, or any linear combination of the latter. Now the first three shadows 
(that is, $j=i,ii,iii$) just correspond to the infinitesimal action of the three-dimensional group of scaling symmetries (gauge transformations) for Somos-5, namely 
\eqref{ScalingSymmetries} 
%sending $x_{2k}\to A_+x_{2k}$, $x_{2k+1}\to A_-x_{2k+1}$, $x_{n}\to B^nx_n$ 
for arbitrary $A_+,A_-,B\in\C^*$. These scaling symmetries leave $w_n$ 
invariant, and by extending them to $A_+,A_-,B\in\D^*$, at the level of the linearization in the plane this means that they do not appear: substituting $y_n=y_n^{(i)}$, $y_n^{(ii)}$,  $y_n^{(iii)}$ 
or any linear combination thereof into (\ref{vxy}) gives a trivial shadow sequence in the plane, namely $(v_n,v_{n+1})=(0,0)$ for all $n$. The fourth shadow sequence (or multiples of it) corresponds 
to a sequence of points on the tangent line to the level curve of the first integral $J^{(0)}$ in the $(w_1,w_2)$ plane: substituting $y_n=y_n^{(iv)}$ into (\ref{vxy}) gives a sequence 
of points $(v_n,v_{n+1})$ that satisfy 
\beq\label{tgtv}
\left(1-\frac{\al^{(0)}}{w_n^2} -\frac{\be^{(0)}}{w_n^2w_{n+1}} \right)\, v_n 
+ 
 \left(1-\frac{\al^{(0)}}{w_{n+1}^2} -\frac{\be^{(0)}}{w_nw_{n+1}^2} \right)\, v_{n+1} = 0,  
%+ \al^{(1)} \left(\frac{1}{w_1}+\frac{1}{w_2}\right) + \frac{\be^{(1)}}{w_1w_2}
\eeq 
which corresponds to $J^{(1)}=0$ (with $\al^{(1)}=0$, $\be^{(1)}=0$ as before). Each generic level curve of $J^{(0)}$ 
has genus 1, being isomorphic to a Weierstrass 
elliptic curve  (\ref{Somos5Curve}) as in Theorem \ref{SomosSolutionTheorem}, and the vector $(v_n,v_{n+1})$ satisfying (\ref{tgtv}) 
is tangent to the level curve at the point $(w_n,w_{n+1})$. 
In Figures 1 \& 2, an orbit of these tangent vectors (corresponding to the original Somos-5 sequence) can be seen to lie on a closed curve (shown in red) starting from the point 
$(v_1,v_2)=(-1,1)$.  
The fifth shadow sequence corresponds to linear perturbations of the QRT map that are transverse to the level curves of the first integral   $J^{(0)}$, involving modular 
deformations of the underlying elliptic curve:   substituting $y_n=y_n^{(v)}$  into (\ref{vxy}), or taking this shadow sequence with a linear combination of any of the other four, gives  
points $(v_n,v_{n+1})$ that produce a non-zero value of the second invariant   $J^{(1)}$ in (\ref{jints}). Figure 1 shows 1000 points on an orbit starting from 
$(v_1,v_2)=(0,1)$, which gives $J^{(1)}=-1$, and Figure 2 shows the same number of points on an orbit starting from 
$(v_1,v_2)=(1,1)$, which gives $J^{(1)}=-2$. Both of these orbits that include transverse perturbations appear to spiral gradually out from the origin.

\section{Conclusion}

We have shown that the main results on the dual Somos-4 sequences from \cite{hone2021casting} have exact analogues for dual Somos-5 sequences: analytic solutions 
in terms of Weierstrass functions with arguments in $\D$; an algebraic form of the general solution for the odd parts using variation of parameters; 
and exact algebraic expressions in terms 
of $\D$-valued Hankel determinants. We have also constructed explicit formulae for a complete basis of shadow Somos-5 sequences, in the sense of 
 \cite{ovsienko2021shadow}, and showed how to reproduce other examples of dual sequences studied by Ovsienko and Tabachnikov. 

The fact that we are able to obtain so many explicit results is due to integrability  lurking in the background, which persists when we move from the   shadows 
to dual  sequemces. From a naive viewpoint, integrability  just means that the original system has a certain number of conserved quantities, 
and these  extend in the obvious way when all variables and parameters are replaced by dual numbers. However, there appears to be a much stronger set of properties 
 that is inherited in the context of dual numbers, since we have shown that the symplectic structure for the QRT map associated with Somos-5 extends to give a bi-Hamiltonian 
structure in the dual setting, and this leads to a proof that the dual QRT map is integrable in the Liouville sense.

Supersymmetric analogues of integrable PDEs and the construction of their associated  bi-Hamiltonian structures have been considered for several decades \cite{kup,jose, op}, and 
supersymmetric soliton equations continue to be a subject of current interest \cite{lou}. However, some analogues of discrete integrable systems over Grassmann algebras have only 
been obtained quite recently \cite{giota, giota2, kk}, and there does not appear to be a fully developed theory of Liouville integrability in this context.       

%SUSY integrable PDEs: 
%discrete SUSY maps:  

\vspace{.1in}
\noindent \textbf{Acknowledgments:} This research was supported by 
the  grant 
 IEC\textbackslash R3\textbackslash 193024 from the Royal Society, and we are both grateful to Ralph Willox and Takafumi Mase for hospitality when we visited the Graduate School 
of Mathematical Sciences, University of  Tokyo in 2023. ANWH also thanks ICMS and the organisers of ISLAND 6 for supporting his attendance at the meeting on Skye.  
%\bibliographystyle{unsrt}
%\bibliography{References}

\end{document}